\documentclass[10pt]{iopart}

\usepackage{iopams}
\usepackage{float}
\usepackage{xspace}
\usepackage{graphicx}
\usepackage{dcolumn}
\usepackage{bm}
\usepackage{color}
\usepackage{soul}
\usepackage{multirow}
\usepackage{hyperref}
\hypersetup{
    unicode=false,          
    pdftoolbar=true,        
    pdfmenubar=true,        
    pdffitwindow=false,     
    pdfstartview={FitH},    
    pdftitle={My title},    
    pdfauthor={Author},     
    pdfsubject={Subject},   
    pdfcreator={Creator},   
    pdfproducer={Producer}, 
    pdfkeywords={keyword1} {key2} {key3}, 
    pdfnewwindow=true,      
    colorlinks=true,       
    linkcolor=blue,          
    citecolor=blue,        
    filecolor=blue,      
    urlcolor=blue,       
    plainpages=false        
}

\newcommand{\TSDW}{\ensuremath{T_\mathrm{SDW}}\xspace}
\newcommand{\Ts}{\ensuremath{T_\mathrm{S}}\xspace}

\newcommand{\Tc}{\ensuremath{T_\mathrm{c}}\xspace}
\newcommand{\BFCA}{\mbox{Ba(Fe$_{\mathrm{1-x}}$Co$_{\mathrm{x}}$)$_{2}$As$_2$}\xspace}

\newcommand{\BKFA}{\mbox{Ba$_{\mathrm{1-x}}$K$_\mathrm{x}$Fe$_{2}$As$_2$}\xspace}
\newcommand{\BFA}{\mbox{BaFe$_{2}$As$_2$}\xspace}
\newcommand{\NFCA}{\mbox{Na(Fe$_{\mathrm{1-x}}$Co$_{\mathrm{x}}$)As$_2$}\xspace}

\newcommand{\Alg}{\texorpdfstring{\ensuremath{A_{1g}}\xspace}{A1g}}

\newcommand{\Blg}{\texorpdfstring{\ensuremath{B_{1g}}\xspace}{B1g}}
\newcommand{\BZg}{\texorpdfstring{\ensuremath{B_{2g}}\xspace}{B2g}}

\newcommand{\Eg}{\texorpdfstring{\ensuremath{E_{g}}\xspace}{Eg}}

\begin{document}

\title[Raman scattering in Fe-based systems]{Fluctuations and pairing in Fe-based superconductors: Light scattering experiments}

\author{N Lazarevi\'c}
\address{Center for Solid State Physics and New Materials, Institute of Physics Belgrade, University of Belgrade, Pregrevica 118, 11080 Belgrade, Serbia}
\ead{nenadl@ipb.ac.rs}
\author{R Hackl}
\address{Walther Meissner Institut, Bayerische Akademie der Wissenschaften, 85748 Garching, Germany}
\ead{hackl@wmi.badw.de}
\vspace{10pt}
\begin{indented}
\item[]\today
\end{indented}

\begin{abstract}
  Inelastic scattering of visible light (Raman effect) offers a window into properties of correlated metals such as spin, electron and lattice dynamics as well as their mutual interactions. In this review we focus on electronic and spin excitations in Fe-based pnictides and chalcogenides in particular, but not exclusively superconductors. After a general introduction to the basic theory including the selection rules for the various scattering processes we provide an overview over the major results. In the superconducting state below the transition temperature \Tc the pair-breaking effect can be observed, and the energy gap can be derived. The energies can be associated with the gaps and their anisotropy on the electron and hole bands. In spite of the similarities of the overall band structures the results are strongly dependent on the family and may even change qualitatively within one family. In some of the compounds strong collective modes appear below \Tc. In Ba$_{1-x}$K$_x$Fe$_2$As$_2$, which has the most isotropic gap of all Fe-based superconductors, there are indications that these modes are exciton-like states appearing in the presence of a hierarchy of pairing tendencies. The strong in-gap modes observed in Co-doped NaFeAs are interpreted in terms of quadrupolar orbital excitations which become undamped in the superconducting state. The doping dependence of the scattering intensity in Ba(Fe$_{1-x}$Co$_x$)$_2$As$_2$ is associated with a nematic resonance above a quantum critical point and interpreted in terms of a critical enhancement at the maximal \Tc. In the normal state the response from particle-hole excitations reflects the resistivity. In addition, there are contributions from presumably critical fluctuations in the energy range of $k_{\rm B}T$ which can be compared to the elastic properties. Currently it is not settled whether the fluctuations observed by light scattering are related to spin or charge. Another controversy relates to possible two-magnon excitations, typically in the energy range of 0.5\,eV. Whereas this response can also originate from charge excitations in most of the Fe-based compounds theory and experiment suggest that the excitations in the range 60\,meV in FeSe are from localized spins in a nearly frustrated system.

\end{abstract}

%
%
%
\maketitle
%
\ioptwocol

\section{Introduction}
\begin{figure}[h]
  \centering
  \includegraphics[width=85mm]{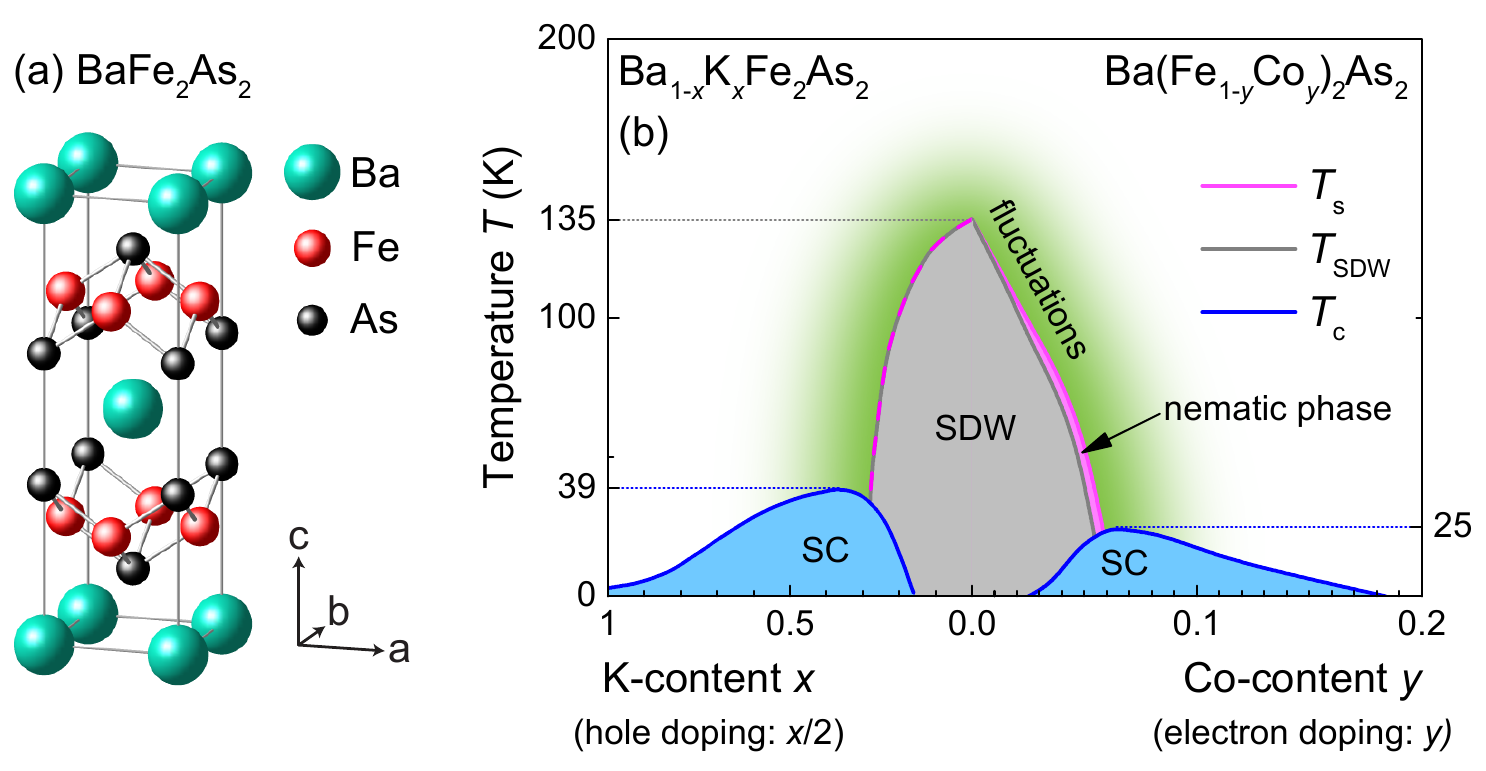}
  \caption[]{ Crystal structure and phase diagram. (a) Crystal structure of BaFe$_2$As$_2$. Thin grey lines indicate the edges of the unit cell (2 Fe per layer). Grey connecting lines between Fe and As illustrate covalent Fe-As bonds. (b) Phase diagram. The spin density wave (SDW) and the superconducting (SC) ranges are indicated in grey and blue, respectively. The dashed pink and grey line indicates a simultaneous structural transition at \Ts and SDW transition at \TSDW. The green shaded area indicates the existence of fluctuations next to the SDW transition. Note that the scales in $x$ (hole doping) and $y$ (electron doping) differ. From \cite{Bohm:2017d} with permission.}
  \label{fig:structure}
\end{figure}

Superconductivity in iron-based componds (FeBCs) was shocking when first reported by Kamihara and coworkers \cite{Kamihara:2006,Kamihara:2008}. The FeBCs consist of quadratically coordinated Fe planes sandwiched between layers of pnictogen (As, P) or chalcogen (S, Se, Te) atoms as shown in Fig.~\ref{fig:structure}\,(a). The rest of the structure is rather variable as can be seen from the sum formulae in Table~\ref{tab:mat}. For this variability and the relative change in the band strucures the FeBCs are a laboratory for studying the interrelation of magnetism, fluctuations and superconductivity or strong \textit{versus} weak-coupling effects as summarized in excellent reviews including \cite{Paglione:2010,Johnston:2010,Stewart:2011,Hirschfeld:2011,Hirschfeld:2016n,Korshunov:2018}.

FeBCs typically have a magnetically ordered phase at zero nominal doping. Upon elemental substitution or application of pressure, magnetism can be suppressed and superconductivity (SC) may appear [Fig.~\ref{fig:structure}\,(b)]. In contrast to the cuprates, all phases are metallic. The order at zero doping is a stripe-like antiferromagnetic spin-density wave (SDW) and widely believed to originate from the nesting properties of the hole- and electron-like Fermi surfaces encircling the $(0,0)$ ($\Gamma$) and the $(\pm\pi,0)/(0,\pm\pi)$ ($X/Y$) points in the idealized 1\,Fe Brillouin zone. Similarly, the topology of the Fermi surface is considered important for superconductivity \cite{Mazin:2008}. Although the Fermi surfaces are always centered at $\Gamma$ and $X/Y$ there are substantial variations in shape and character across the families and as a function of doping \cite{Kordyuk:2013}. However, this variation is not necessarily at the origin of the differences in the superconducting states of the FeBC. Already on the basis of the band structure on the level of local-density approximation (LDA) \cite{Graser:2009}, nearly degenerate superconducting ground states can be derived \cite{Hirschfeld:2011,Thomale:2011}.

In addition to SDW and SC phase transitions, nematic order, with the rotational symmetry broken but the translational symmetry preserved, and wide temperature ranges with fluctuation are observed \cite{Fernandes:2014}. For studying this plethora of instabilities, a wide variety of experimental methods has been applied. Inelastic light scattering a useful technique, since relevant information on practically all phases and their fluctuations can be obtained.

It will be the purpose of this review to present typical results, provide a snapshot of the current status of the field and outline possible future developments. Firstly, we briefly describe the experiment and the theoretical background and then summarize the most relevant results obtained from light scattering with the focus placed on electronic and spin excitations.

\section{Raman experiment}
\label{sec:exp}
\begin{figure}
  \centering
  \includegraphics [width=8.5cm]{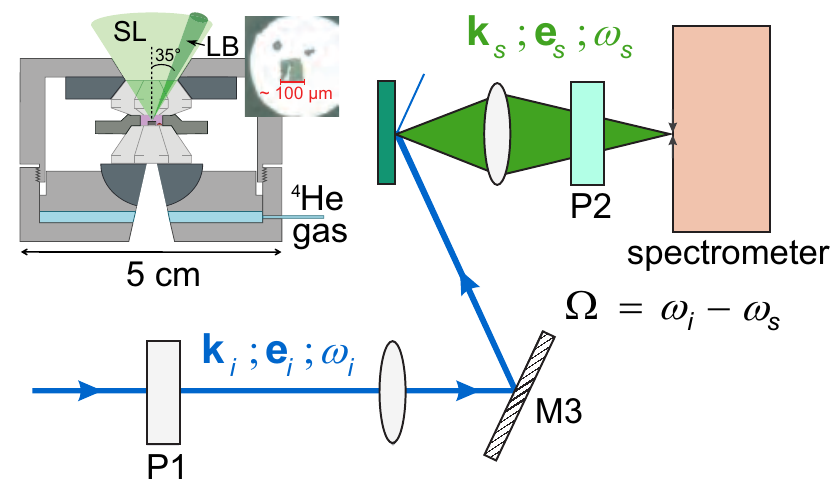}
  \caption[]{Schematic representation of a Raman experiment. The polarized monochromatic incident photons hit the sample at a large angle of incidence. The scattered photons are collected along the surface normal. Before entering the spectrometer the scattered photons pass an analyzer. Inset: Side view of a Raman pressure cell. The laser beam (LB) enters from the right, the scattered light (SL) is collected along the normal of the sample surface. From \cite{Massat:2018}.
  }
  \label{fig:exp}
\end{figure}

Shown in Fig.~\ref{fig:exp} is a schematic view of the experimental setup of a typical macro Raman experiment \textit{in vacuo} and in a diamond anvil cell (inset) on opaque samples. In the macro setup the incident light with polarization $\hat{\bf e}_I$ impinges on the surface at an angle of incidence $\vartheta_I\sim 70^{\rm o}$ in order to prevent the directly reflected light to enter the optics and the spectrometer. For this ``pseudo-Brewster'' angle the reflection is minimal for $\hat{\bf e}_I$  parallel to the plane of incidence. The scattered light is collected along the surface normal. Photons having a selected polarization state $\hat{\bf e}_S$ enter the spectrometer.
A charge-coupled device (CCD) detector registeres the number of transmitted photons per unit time $\dot{N}_{I,S}(\Omega)$ (``Raman spectrum'') for a given energy shift $\Omega = \omega_I-\omega_S$ and polarization combination ($\hat{\bf e}_I,\hat{\bf e}_S$), where $\omega_{I,S}$ is the energy of the photons. From linear combinations of the measured spectra, pure symmetries $\mu$ can be derived (for details see Section \ref{sec:selection rules}). The differential light scattering cross-section is proportional to the Raman spectrum,
\begin{equation}
  \frac{d^2\sigma_{I,S}}{d\Omega_S d\omega_S} = \hbar r_0^2\frac{\omega_S}{\omega_I}\frac{1}{\pi}\{1+n(\Omega, T)\}R_{I,S}\chi^{\prime \prime}_{I,S}({\bf q}, \Omega, T).
  \label{eq:sigma}
\end{equation}
Here $\Omega_S$ is the solid angle into which the photons are scattered, and $R_{I,S}$ absorbs matrix element effects and experimental factors, $\chi_{\mu}({\bf q}, \Omega, T) = \chi^{\prime}_{\mu}+i\chi^{\prime \prime}_{\mu}$ is the typically non-resonant response function, $n(\Omega, T) = [\exp{\hbar\Omega/k_BT}-1]^{-1}$ is the Bose-Einstein occupation number and $r_0=e^2/(4\pi\varepsilon_0 mc^2)$ the Thompson electron radius, thus Eq.~(\ref{eq:sigma}) describes the cross section per electron.

\section{Materials}
\label{sec:mat}
Most of the existing FeBCs were studied by Raman scattering. In the beginning the phonons were in the main focus \cite{Hadjiev:2008,Litvinchuk:2008,Rahlenbeck:2009,Um:2012,Lazarevic:2011,Lazarevic:2012,ZhangQM:2012}. With the advent of high-quality single crystals of the 122 family \cite{Rotter:2008}, being a result of the FeAs self-flux growth \cite{Sefat:2008}, the study of electronic properties by light scattering became promising, and the superconducting gap was studied successfully in \BFCA for two doping levels \cite{Muschler:2009}. Soon thereafter the redistribution of spectral weight in the SDW state of \BFA was reported \cite{Chauviere:2011,Sugai:2012}. Finally, the fluctuations above the magneto-structural transformation, inferred from transport \cite{Chu:2010}, were observed \cite{Choi:2010,Gallais:2013}. Although there are reports on spectra in the range 1\,000\,cm$^{-1}$ to 4\,000\,cm$^{-1}$ (the energy range of two-magnon excitations) \cite{Sugai:2011}, their interpretation remains controversial.

The materials for which data on the spin, charge and orbital response exist are compiled in Table \ref{tab:mat}.
\begin{table*}
  \caption{\label{tab:mat}Materials studied by Raman scattering. The table includes only materials for which electronic properties (spin, charge, fluctuations) were studied. In the second column typical acronyms are listed which will be used occasionally in the text. The main subject of the respective experiments are listed in the second-last column.
  }
  \centering
  \footnotesize
\begin{tabular}{@{}lllll}
  \br
  material&acronym&substitution&subject&reference\\
  \mr
  BaFe$_2$As$_2$&BFA/Ba122&no&SDW&\cite{Sugai:2012,Sugai:2013}\\
  SrFe$_2$As$_2$&Sr122&no&SDW, fluctuations&\cite{Choi:2010,W.L.Zhang:2014}\\
  EuFe$_2$As$_2$& Eu122&no&SDW,fluctuation&\cite{W.L.Zhang:2014,W.L.Zhang:2016}\\
  CaFe$_2$As$_2$& Ca122&no&SDW&\cite{W.L.Zhang:2016}\\
  \mr
  Ba(Fe$_{1-x}$Co$_{x}$)$_{2}$As$_{2}$&BFCA&$x=0.061, 0.085$&SC gap&\cite{Muschler:2009}\\
  Ba(Fe$_{1-x}$Co$_{x}$)$_{2}$As$_{2}$&BFCA&$x=0.061$&gap, vertex&\cite{Mazin:2010a}\\
  Ba(Fe$_{1-x}$Co$_{x}$)$_{2}$As$_{2}$&BFCA&$x=0.08$&gap&\cite{Sugai:2010}\\
  Ba(Fe$_{1-x}$Co$_{x}$)$_{2}$As$_{2}$&BFCA&$0\le x \le 0.045$&SDW&\cite{Chauviere:2011}\\
  Ba(Fe$_{1-x}$Co$_{x}$)$_{2}$As$_{2}$&BFCA&$0\le x \le 0.20$&fluctuations&\cite{Gallais:2013}\\
  Ba(Fe$_{1-x}$Co$_{x}$)$_{2}$As$_{2}$&BFCA&$0.055\le x \le 0.10$&SC gap&\cite{Chauviere:2010}\\
  Ba(Fe$_{1-x}$Co$_{x}$)$_{2}$As$_{2}$&BFCA&$0\le x \le 0.10$&SC gap-nematicity&\cite{Gallais:2016}\\
  Ba(Fe$_{1-x}$Co$_{x}$)$_{2}$As$_{2}$&BFCA&$x=0, 0.025, 0.051$&fluctuations&\cite{Kretzschmar:2016}\\
  Ba(Fe$_{1-x}$Co$_{x}$)$_{2}$As$_{2}$&BFCA&$0.045\le x \le 0.085$&fluctuations and SC gap&\cite{Bohm:2018}\\
  Ca(Fe$_{1-x}$Co$_{x}$)$_{2}$As$_{2}$& &$x=0.03$&crystal field&\cite{Kumar:2014}\\
  Sr(Fe$_{1-x}$Co$_{x}$)$_{2}$As$_{2}$& &$x=0, 0.04$&SDW&\cite{Y.X.Yang:2014}\\
  Ba(Fe$_{1-x}$Au$_{x}$)$_{2}$As$_{2}$& &$x=0, 0.012, 0.014, 0.031$ & fluctuations& \cite{Wu:2017origin}\\
  \mr
  Ba$_{1-x}$K$_x$Fe$_2$As$_2$&BKFA&$x=0.4$&SC gap and pairing&\cite{Kretzschmar:2013}\\
  Ba$_{1-x}$K$_x$Fe$_2$As$_2$&BKFA&$x=0.4$&SC gap and pairing&\cite{Bohm:2014}\\
  Ba$_{1-x}$K$_x$Fe$_2$As$_2$&BKFA&$0.22\le x \le 0.70$&fluctuations and SC gap&\cite{Bohm:2018,Bohm:2018}\\
  Ba$_{1-x}$K$_x$Fe$_2$As$_2$&BKFA&$x=0.25, 0.4, 0.6$&fluctuations and SC gap&\cite{WuSF:2017}\\
  \mr
  BaFe$_2$(As$_{1-x}$P$_{x}$)$_2$&BFAP&$x=0.5$&fluctuations and SC gap&\cite{WuSF:2017}\\
  \mr
  NaFe$_{1-x}$Co$_{x}$As&Na111&$0 \le x \le 0.08$&fluctuations and SC gap&\cite{Thorsmolle:2016}\\
  \mr
  CaKFe$_4$As$_4$ &CKFA& no & SC gap and pairing & \cite{Jost:2018,ZhangWL:2018} \\
  \mr
  Fe$_{1+\delta}$Te$_{1-x}$Se$_x$& &$x=0, 0.4$&SC gap, phonon, magnon&\cite{Okazaki:2011}\\
  FeSe&11&no&fluctuations&\cite{Massat:2018,Massat:2016,Glamazda:2019}\\
  FeSe&11&no&2-magnon&\cite{Baum:2019}\\
  FeSe$_{0.82}$&11&no&crystal field&\cite{Kumar:2010}\\
  FeSe$_{1-x}$S$_x$ & & $x=0, 0.04, 0.08, 0.15, 0.20$ & fluctuations& \cite{Zhang:2017s}\\
  \mr
  K$_{0.75}$Fe$_{1.75}$Se$_{2}$& &no&gap&\cite{Khodas:2014}\\
  Rb$_{0.8}$Fe$_{1.6}$Se$_{2}$& &no&gap&\cite{Kretzschmar:2013}\\
  $A_{0.8}$Fe$_{1.6}$Se$_{2}$& &$A=$\,K, Rb, Cs, Tl&2-magnon& \cite{A.M.Zhang:2012}\\
  \br
\end{tabular}\\

\end{table*}
\normalsize
\section{Theoretical background}
\label{sec:theory}
The analysis of the results in the FeBCs requires insight into the theoretical background, including both standard knowledge and modern developments. We give a brief historical summary and sketch the underlying theory.

\subsection{Historical remarks}
Raman scattering studies of excitations of localized spins started in the 1960s on insulating antiferromagnets \cite{Fleury:1966} and it experienced a renaissance with the advent of the cuprates \cite{Lyons:1989}. The theoretical framework was set by the seminal work of Fleury and Loudon \cite{Fleury:1968} which still represents the basis of contemporary analyses \cite{Devereaux:2007} even in the case of metallic systems such as the FeSCs \cite{Sugai:2011,Chen:2011b,Baum:2018}.

Light scattering from conduction electrons was first discussed in the context of superconductors \cite{Abrikosov:1961}. It took almost 20 years to observe the effect experimentally in the layered compound $2H$-NbSe$_2$ \cite{Sooryakumar:1980}. In NbSe$_2$ SC competes with a charge density wave (CDW) for the area on the Fermi surface (FS), and the spectral features observed below $T_c$ cannot directly be traced back to the superconducting energy gap such as for selected symmetries in the A15 compounds V$_3$Si and Nb$_3$Sn \cite{Dierker:1983,Hackl:1983} or in the cuprates \cite{Devereaux:2007,Cooper:1988a,Hackl:1988}. Hence, only in special cases the Raman spectra of superconductors can be described satisfactorily in terms of lowest order weak coupling theory as developed between 1961 and 1984 \cite{Abrikosov:1961,Abrikosov:1973,Klein:1984}. In all other cases, including the FeBCs, lowest order is insufficient, although it still captures the plain vanilla such as the strong momentum dependence of the gap if the symmetries of the response are properly taken into account \cite{Devereaux:1994}.

In normal metals, contributions from particle-hole excitations were observed and discussed for doped semiconductors \cite{Chandrasekhar:1977,Ipatova:1981}, but in-depth studies started only in the cuprates \cite{Zawadowski:1990,Staufer:1990,Slakey:1991,Hackl:1996,Opel:2000}. Similarly as in the superconducting state, the major contribution from Raman scattering, in addition to what was known from optical spectroscopy, was the observation of a polarization dependent relaxation of the carriers which could be mapped on the electronic momentum \cite{Devereaux:1994,Einzel:1996}.

Important new developments pertain to the inclusion of an anisotropic pairing potential in the superconducting state  \cite{Klein:1984,Zawadowski:1972,Monien:1990,Chubukov:2009,Scalapino:2009,Maiti:2016}, amplitude (``Higgs'') fluctuations of the superconducting order parameter \cite{Littlewood:1981,Littlewood:1982}, number-phase fluctuations in multiband systems (Leggett modes \cite{Leggett:1966}) \cite{Blumberg:2007,Klein:2010,Burnell:2010,Cea:2016,Huang:2018} or a nematic resonance \cite{Gallais:2016}. Whereas the interpretation of the results in NbSe$_2$ in terms of coupled gap excitations and amplitude fluctuations in a coupled SC-CDW system seems to converge \cite{Measson:2014,Pekker:2015} the discussion of the \Eg symmetry contributions in the A15 materials \cite{Dierker:1983,Hackl:1983,Hackl:1989,Monien:1990,Pekker:2015}, the \Blg response in the cuprates \cite{Kendziora:1995,Chen:1997,Sugai:2000,LeTacon:2006,Munnikes:2011,LiY:2013} or the in-gap modes in the FeBCs \cite{Gallais:2016,Bohm:2018,Kretzschmar:2013,Bohm:2014,Thorsmolle:2016} remains controversial. In addition to the weak-coupling description of the superconducting state at $T=0$ a lot more work is needed to arrive at a coherent picture for the normal and superconducting states in the presence of collisions and strong coupling. Only a few special cases have been studied theoretically so far \cite{Zawadowski:1990,Devereaux:1992,Devereaux:1993,Devereaux:1995,Manske:2004,Caprara:2005,%
Caprara:2015,Karahasanovic:2015,Gallais:2016a}.

More details can be found in Refs.~\cite{Devereaux:2007} and \cite{Einzel:1996} and, for recent developments, in \cite{Bohm:2018,Chen:2011b,Baum:2018,Scalapino:2009,Maiti:2016,Cea:2016,%
Huang:2018,Karahasanovic:2015,Gallais:2016a,Boyd:2009,Khodas:2015}.

\subsection{Light scattering}
\label{sec:LS}

Photons do not directly scatter off low energy excitations. Rather, high-energy electron-hole pairs having energies of the incoming photons $\hbar\omega_I$ are created and couple to excitations in the range $\hbar\Omega = {\cal O}(k_BT)$ such as phonons, fluctuations, particle-hole, gap or spin excitations. After scattering, the electron-hole pairs recombine and emit photons with energies $\hbar\omega_S = \hbar\omega_I \mp \hbar\Omega$. The mechanism works whether or not the intermediate electronic states are eigenstates. If they are eigenstates the cross-section is resonantly enhanced but most of the results were successfully analysed in terms of non-resonant scattering. In systems with a high correlation energy $U = {\cal O}(\hbar\omega_I)$ there are no well-defined eigenstates, and the resonances are in fact found to be mild in most of the cases.

Given these considerations, Eq.~(\ref{eq:sigma}) is most naturally derived from scattering matrix elements in third order perturbation theory. For many practical purposes the Raman susceptibility or response function $\chi(\Omega, T)$ is derived in the non-resonant limit by various techniques. 

\begin{figure}
  \centering
  \includegraphics [width=8cm]{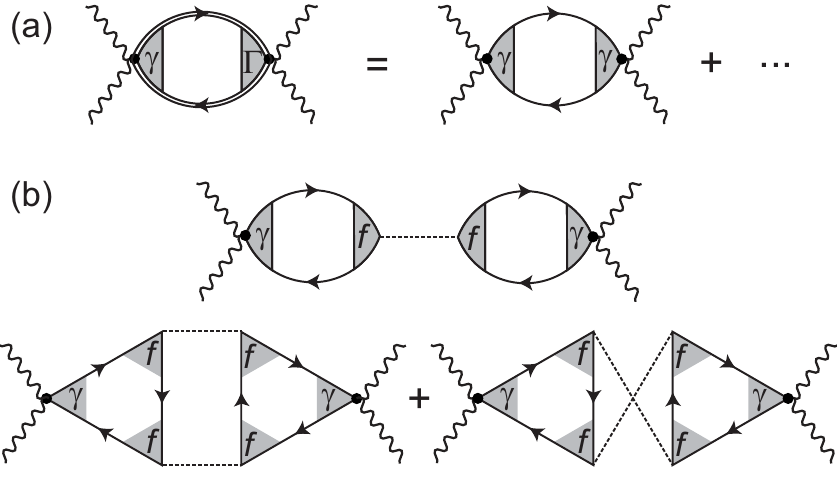}
  \caption[]{Feynman diagrams for (non-resonant) light scattering. (a) Raman response of particle-hole excitations in the presence of interactions and (b) scattering processes involving one and two fluctuations. Wavy lines represent incident and scattered photons whereas solid lines are electronic propagators. The bosonic fluctuation propagators are represented by dashed lines. The bare and renormalized Raman vertices $\gamma$ and, respectively, $\Gamma$ describe the interaction of light and electrons and $f$ describes the interaction of electrons and other excitations, e.g. bosons.
  }
  \label{fig:feynman}
\end{figure}

For better visualization, Feynman diagrams for Raman scattering on charge or spin excitations are displayed in Figure~\ref{fig:feynman}. $\gamma$ and $\Gamma$ are the bare and the renormalized vertices, respectively, and depend on the light polarizations and on momentum. In this way they determine the selection rules.

\subsection{Charged systems}
\label{sec:LS-metals}
Metals have a small penetration depth for light $a < \delta_{0} < \lambda_{I,S}$ where $a$ is a typical lattice constant. As a consequence, the momentum transfer in a metal is bigger than in an insulator, $|{\bf q}| = q \sim \delta_{0}^{-1}$, where 100\,\AA\,$ < \delta_{0} < $\,1000\,\AA\, for typical metals, as pointed out first by Abrikosov and Fal'kovskii \cite{Abrikosov:1961}. Yet, the available momentum is still much smaller than $\pi/a$, and for exciting a non-interacting conduction electron from an occupied (i) to an unoccupied (f) state the maximal energy is limited by $\hbar\Omega_{\rm max} = \epsilon_{\bf k}^{(f)}-\epsilon_{\bf k}^{(i)} \approx \hbar v_F\delta_0^{-1}$ where $\epsilon_{\bf k}$ is the electronic dispersion. In the FeBCs the relevant in-plane Fermi velocity is of the order $10^{6}$\,cm/s, and the penetration depth is close to 1000\,\AA\, yielding $\hbar\Omega_{\rm max} \lesssim 1\,{\rm meV}\equiv 8\,{\rm cm^{-1}}$.

Due to charge conservation and screening all isotropic charge excitations are pushed up to the plasma frequency $\Omega_{\rm pl}$ \cite{Pines:1966}, and without scattering in the range of $k_BT$ in non-interacting systems with a strictly quadratic electron dispersion. What looks like a disadvantage at first glance is the origin of the selection rules for electronic Raman scattering \cite{Abrikosov:1973}. Loosely speaking, one cannot move charges from one unit cell to another one but {the charges can only be redistributed in phase inside all unit cells} (quadrupolar-type of excitations). This is the origin of the form factors, and the light does not scatter from the charge density but from a weighted charge density,
\begin{equation}
  \label{eq:Rdensity}
  \tilde{\rho}_{\bf q}=\frac{1}{N}\sum_n\sum_{{\bf k},\sigma}\gamma_n({\bf k},{\bf q})c^\dagger_{n,{\bf k}+{\bf q},\sigma}c_{n,{\bf k},\sigma}.
\end{equation}
$n$ is the band index, and $\gamma_n({\bf k},{\bf q})$ is a form factor which is related to the Raman vertex $\gamma_{\alpha,\beta}({\bf k},{\bf q})$ through the polarization directions $\hat{\bf e}_{I,S}$,
\begin{equation}
  \label{eq:Rvertex}
  \gamma_n({\bf k},{\bf q})=\sum_{\alpha,\beta}e_I^\alpha\gamma_{n,\alpha,\beta}({\bf k},{\bf q})e_S^\beta.
\end{equation}
The bare response $\tilde{\chi}({\bf q},\Omega)$ is the commutator of $\tilde{\rho}_{\bf q}$,
\begin{equation}
  \label{eq:chibare}
  \tilde{\chi}({\bf q},\Omega) = \langle\langle[\tilde{\rho}_{\bf q}(t),\tilde{\rho}_{-\bf q}(0)]\rangle\rangle_\Omega
  = \tilde{\chi}^\prime + i\tilde{\chi}^{\prime\prime},
\end{equation}
where $\langle\langle\dots \rangle\rangle_\Omega$ denotes the thermodynamic average and the Fourier transformation, and $\tilde{\chi}^\prime$ and $\tilde{\chi}^{\prime\prime}$ are the real and imaginary part of $\tilde{\chi}$, respectively. Eq.~(\ref{eq:chibare}) includes a sum over the Brillouin zone and the bands through Eq.~(\ref{eq:Rdensity}), and can be recast as,
\begin{equation}
  \label{eq:chibare2}
  \tilde{\chi}_{a,b}({\bf q},\Omega) = \frac{1}{N}\sum_n\sum_{{\bf k}}a_{\bf k}b_{\bf k}\Theta_{n}({\bf k},\Omega).
\end{equation}
$a=a_{\bf k}$ and $b=b_{\bf k}$ are generalized vertices that stand for either an isotropic (1) or Raman ($\gamma({\bf k},{\bf q})$) or renormalized Raman ($\Gamma({\bf k},{\bf q})$) vertex. $\Theta_{n}({\bf k},\Omega)$ is the response kernel, examples of which will be presented below.

The bare response is not gauge invariant, and charge conservation leads to the final - exact - result for the response \cite{Monien:1990,Devereaux:1995a,Maiti:2017},
\begin{equation}
  \label{eq:chisc}
  \chi_{\gamma,\gamma} = \tilde{\chi}_{\gamma,\gamma} - \frac{\tilde{\chi}_{1,\gamma}\tilde{\chi}_{\gamma,1}}{\tilde{\chi}_{1,1}} \left(1-\frac{1}{\varepsilon}\right).
\end{equation}
The vertex is written down explicitly as a subscript in Eq.~(\ref{eq:chisc}), and $\varepsilon$ is the dielectric function. Since a constant vertex can be pulled in front of the sum of Eq.~(\ref{eq:Rdensity}) the first two terms of Eq.~(\ref{eq:chisc}) cancel for constant $\gamma$ (corresponding to lowest order $A_{1g}$), and only the last term survives but is suppressed as $q^2/\Omega_{\rm pl}^2$ in a charged system. In the fully symmetric channel, having \Alg symmetry in the $D_{4h}$ space group, applicable for most of the FeBCs, the response is at least partially screened.

\subsection{Weakly interacting systems}
\label{sec:weak}
The scattering from free electrons cuts off at $\hbar\Omega_{\rm max}$ \cite{Platzman:1965}. In a realistic normal metal with a single conduction band either impurities or interactions can maintain the $q=0$ selection rule and facilitate the occurrence of broad continua  extending to energies well above $\hbar\Omega_{\rm max}$ \cite{Zawadowski:1990,Kostur:1992}. Obviously this is the case in all materials of interest here including the FeBCs, the cuprates, the A15 compounds, MgB$_2$ and many others. However, only in the cuprates the argumentation is straightforward since in the single-layer compounds with only one CuO$_2$ plane per unit cell, for instance ${\rm La_{2-x}Sr_{x}Cu_{4}}$, there is only one conduction band. Somewhat surprisingly, the response in the double-layer compounds (two neighboring CuO$_2$ planes) such as ${\rm YBa_{2}Cu_{3}O_{6+y}}$ is nearly identical to that in materials with just one CuO$_2$ plane. It was in fact shown that the contributions from individual bands in multi-band systems just add up to a good approximation not only in the cuprates \cite{Devereaux:1996} but also in the FeBC \cite{Boyd:2009} allowing one, as a starting point, to treat the response in a band basis and neglect inter-band transition.

If we disregard spurious contributions to the cross section such as luminescence (being a good approximation at low energies; see, however, Ref. \cite{Zhang:2017s}) only (dynamic) interactions can produce the broad continua observed \cite{Opel:2000,Muschler:2010a}. Impurities are present but irrelevant in high-quality single crystals as can be seen directly in a superconductor since impurities reduce the effect of pair breaking in the Raman spectra. As shown by Klein and Dierker \cite{Klein:1984} the response of a clean isotropic superconductor has a square root singularity at the gap edge $2\Delta$ in the limit $q=0$. In the presence of impurities (and similarly for finite $q$ \cite{Klein:1984}) the singularity disappears and the response at $2\Delta$ becomes finite and scales as $\Delta\tau_0$, where $\tau_0$ is the impurity scattering time \cite{Devereaux:1992,Devereaux:1993,Devereaux:1995}. In all realistic cases the difference between the normal and superconducting spectra cannot be observed any further for $\hbar\tau_0^{-1} \to \Delta$. Therefore, the continuum in superconductors essentially comes from dynamical interactions between the conduction electrons and other excitations which are gapped out below $T_c$ for $\hbar\Omega \lesssim 2\Delta$. In nearly all superconductors which have been studied by light scattering this conclusion holds. The only exception is the $A_{1g}$ response in A15 compounds where the normal state intensity vanishes 
 \cite{Hackl:1989,Hackl:1988a}.


Although the normal state continuum generally signals the presence of strong interactions, it is impossible to describe the superconducting response in terms of strong-coupling theory so long as the microscopic origin of the relevant interactions is unknown. Only in the case of spin fluctuations both the normal and the superconducting spectra have been modelled microscopically on equal footing \cite{Manske:2004}. In a few cases phenomenological descriptions on the basis of Eliashberg theory were applied \cite{Devereaux:1995,Cuk:2005,Inosov:2007,Prestel:2010}.

In the majority of cases, the weak-coupling result \cite{Klein:1984} is used since it is sufficient to capture the generic properties of the superconducting state. The response kernel is given by the Tsuneto-Maki (TM) function \cite{Tsuneto:1960},
\begin{equation}
  \label{eq:Tsuneto}
  \Theta_{{\bf k},{\rm TM}}^{\prime\prime}(\Omega,T) =\frac{\pi}{2}\frac{|2\Delta_{\bf k}|^2} {\Omega\sqrt{\Omega^2-|2\Delta_{\bf k}|^2}};~~ \Omega > |2\Delta_{\bf k}|.
\end{equation}
It is the result of a superposition of particle excitations across the gap from an occupied into an unoccupied state and pair breaking. Both contributions have a square-root singularity at $|2\Delta_{\bf k}|$ and, because of the coherence factors, add constructively in the case of light scattering and destructively in the case of the (real part of the) optical (IR) conductivity $\sigma^\prime(\Omega)$\cite{Mattis:1958}.

In the case of a charge or spin density wave the functional form of the response is also described by to Eq.~(\ref{eq:Tsuneto}) \cite{Eiter:2013} if the material is insulating below the transition. In a metal the best way of describing the response is the superposition of a normal metallic response and a condensate reminiscent of a two-fluid model \cite{Einzel:1996}. In all metallic cases there is an incomplete redistribution of spectral weight from low to high energies starting immediately at the transition temperature.



\subsection{Collision limited regime}
The simplest type of response in the normal state in systems with vanishingly small $\hbar v_Fq$ results from the presence of (heavy) impurities. Here the electrons change only their momentum but not their energy (Drude model). The gauge-invariant kernel was derived by Zawadowski and Cardona \cite{Zawadowski:1990},
\begin{equation}
  \label{eq:Drude}
  \Theta_{\bf k}^{\prime\prime}(\Omega)=\frac{\pi}{2}\frac{\hbar\Omega\Gamma_{\bf k}^\ast} {(\hbar\Omega)^2+(\Gamma_{\bf k}^\ast)^2},
\end{equation}
where $\Gamma_{\bf k}^\ast = \hbar(\tau_{\bf k}^\ast)^{-1}$. $\tau_{\bf k}^\ast$ is not identical to the  elec\-tronic relaxation time but is renormalized by a (presumably small) channel-dependent vertex correction, similarly as in ordinary transport where the vertex corrections ensure (among other things) that forward scattering does not contribute to the resistivity.

If the electrons scatter from excitations in the energy range of $k_BT$ such as phonons, spin fluctuations or among themselves they transfer both momentum and energy. As a consequence they become dressed quasi-particles, and the relaxation rate depends now on energy, momentum and temperature. Due to these interaction effects the electrons' velocity gets reduced and the mass increases by the same factor $1+\lambda_{\bf k}(\Omega,T)$, and the response transforms to
\begin{equation}
  \label{eq:extended}
  \Theta_{\bf k}^{\prime\prime}(\Omega,T)=\frac{\pi}{2}\frac{\hbar\Omega\Gamma_{\bf k}^\ast(\Omega,T)}
  {(\hbar\Omega[1+\lambda_{\bf k}(\Omega,T)])^2+(\Gamma_{\bf k}^\ast(\Omega,T))^2}.
\end{equation}
The energy and temperature dependent projected parameters $\Gamma_{\bf k}^\ast(\Omega,T)$ and $1+\lambda_{\bf k}(\Omega,T)$ can be derived if $\Theta_{\bf k}^{\prime\prime}(\Omega,T)$ is known for a sufficiently wide energy interval \cite{Opel:2000}. The zero-energy extrapolation value of $\Gamma_0(T) = \Gamma_{\bf k}^\ast(\Omega\to0,T)$ can be compared with ordinary or optical transport, for instance. Shastry and Shraiman  \cite{Shastry:1990} noticed that the relation between the Raman response and the real part of the optical conductivity, $\Theta_{\bf k}^{\prime\prime}(\Omega,T) \propto \Omega\sigma^\prime(\Omega,T)$, is a good approximation in many cases, in particular if the momentum dependence is week (simply because $\sigma^\prime$ is always an average over the entire Fermi surface whereas $\Theta_{\bf k}^{\prime\prime}$ is not (see below).

An expression equivalent to Eq.~(\ref{eq:extended}) can be derived for the superconducting state. Since it is an order of magnitude more complicated and was not used in the case of the FeBC so far we do not reproduce it here. The interested reader can consult Refs.~\cite{Cuk:2005,Prestel:2010}.

All results summarized here are essentially lowest-order response theory. However, the experiments in \BKFA and \NFCA show \cite{Bohm:2018,Kretzschmar:2013,Bohm:2014,Thorsmolle:2016} that higher order corrections may be necessary for the proper interpretation and for extracting relevant information on the pairing in the superconducting state \cite{Bohm:2014,Chubukov:2009,Scalapino:2009,Karahasanovic:2015,Khodas:2015}.

\subsection{Beyond lowest order}
\label{sec:higher-order}
Eq.~(\ref{eq:Tsuneto}) is the lowest-order approximation of the response and is not gauge invariant as already pointed out by Klein and Dierker \cite{Klein:1984}. The controversy as to whether or not the resulting vertex corrections are relevant for the interpretation in the A15 materials is still not settled \cite{Klein:1984,Monien:1990,Littlewood:1982,Hackl:1989,Varma:2002}. A similar narrow in-gap mode as in the $E_g$ response of the A15s was observed recently in BKFA where it is well separated from the pair-breaking peak \cite{Kretzschmar:2013}. A mode in the $d_{x^2-y^2}$ channel (1\,Fe unit cell) with this property was predicted by Scalapino and Devereaux \cite{Scalapino:2009} for a two-band model applicable to the FeBCs. Chubukov and coworkers predicted an $A_{1g}$ mode originating from the same type of mechanism for a different hierarchy of pairing instabilities.

The existence of in-gap modes in a superconductor was first noticed by Bardasis and Schrieffer (BS) \cite{Bardasis:1961}. BS studied the effect of final state interaction in the presence of an anisotropic pairing potential $V_{{\bf k},{\bf k}^\prime}$ and found undamped modes below the gap edge, $\hbar\Omega_L<2\Delta$, which are characterized by quantum numbers $L$ and $M$ corresponding to the expansion of $V_{{\bf k},{\bf k}^\prime}$ into spherical harmonics. These collective excitations, usually called BS modes, are similar to excitons in semiconductors with binding energy $E_{b,L,M} = 2\Delta-\hbar\Omega_{L,M}$. For simplicity they may be refered to as $E_{{\rm BS},\alpha}$ labelled by $\alpha$ in consecutive order. The result was adopted for light scattering in superconductors \cite{Monien:1990} and is formally similar to light scattering from roton pairs in superfluid $^4$He \cite{Greytak:1969,Zawadowski:1972}. The predictions include symmetry selection rules and the dependence of $E_{{\rm BS},\alpha}$ and the strength of the pole $Z_{{\rm BS},\alpha}$ on the relative coupling strength of the sub-leading channels $\alpha>1$ with  respect to the ground state $\alpha=1$, $\lambda_\alpha/\lambda_1$. As pointed out in Refs. \cite{Chubukov:2009} and \cite{Scalapino:2009} the analysis of these excitons would help in clarifying the so far elusive pairing mechanism in the FeBCs.

The functional form of the response additional to lowest order [Eq.~(\ref{eq:Tsuneto})] reads \cite{Scalapino:2009},
\begin{equation}
  \Delta\tilde{\chi}^{\prime\prime}(\Omega)\! =\! \left(\frac{2}{\Omega}\right)^2 \!\!\Im \left\{\frac{\langle \gamma(\mathbf{k}) g(\mathbf{k})  \Delta(\mathbf{k})\bar{P}(\Omega, \mathbf{k})  \rangle^2}
  {\left( \lambda_d^{-1} \!-\! \lambda_s^{-1} \right) \!-\! \langle g^2 \bar{P}(\Omega, \mathbf{k}) \rangle} \right\}
\label{eq:responsecm}
\end{equation}
and was used in Ref. \cite{Bohm:2014} for estimating the value of the sub-leading $d_{x^2-y^2}$ coupling parameter $\lambda_d$ from the electronic Raman spectra of BKFA. Identical expressions with redefined coupling parameters were derived for explaining the nematic resonance \cite{Gallais:2016}. Similarly as in the case of excitons, there may be more than one BS mode in the presence of several sub-leading coupling channels. This possibility was considered recently in a theoretical study \cite{Maiti:2016}.

In a system with more than one band there is an additional contribution to the response from number-phase oscillations between the bands in momentum space comparable to the Josephson effect in real space \cite{Leggett:1966}. It must be included to make the response gauge invariant, as pointed out recently by Cea and Benfatto \cite{Cea:2016}. Usually, for instance in the cases studied theoretically by Suhl and coworkers \cite{Suhl:1959}, which is possibly realized in MgB$_2$, the main contribution to superconductivity comes from coupling in the individual bands having strength $\lambda_{i,i}$. The inter-band coupling $\lambda_{i,j}$ is weaker but leads to an increase of $T_c$ to values above the maximum of the individual bands. It is widely believed that the main contribution to pairing in the FeBC has its origin in inter-band \cite{Mazin:2008} or inter-orbital \cite{Kontani:2010} coupling and that the intra-band pairing is weak. The effect of weak inter-band coupling was investigated already earlier in the context of MgB$_2$ \cite{Blumberg:2007,Klein:2010} whereas strong inter-band coupling was addressed only recently in the context of the FeBCs \cite{Burnell:2010,Cea:2016,Huang:2018}. The symmetry properties of the Leggett modes depend sensitively on the orbital content \cite{Burnell:2010,Huang:2018}.

\begin{figure}
  \centering
  \includegraphics [width=8.5cm]{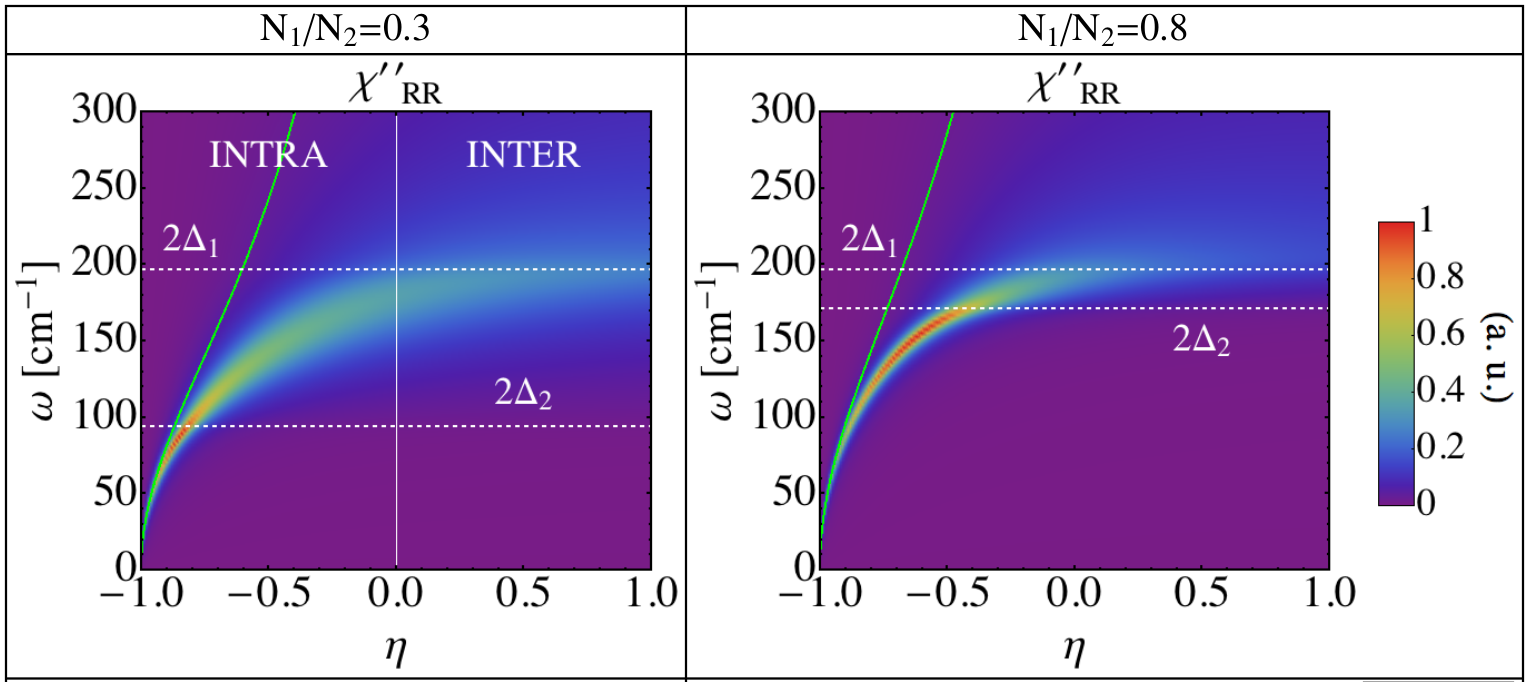}
  \caption[]{ Response from Leggett modes in a two-gap system. $N_i$ is the density of states on band $i$. $\eta$ encodes the ratio of intra- to inter-band coupling with $\eta=-1$ and $\eta=1$ representing pure intra- and inter-band coupling, respectively. The mode's energy saturates at the maximal gap. The damping starts above the smaller gap. From \cite{Cea:2016}.
  }
  \label{fig:Leggett}
\end{figure}
For weak inter-band coupling, the mode related to the number-phase fluctuations is below the gap energy and increases essentially linearly with $\lambda_{i,j}$. For $\lambda_{i,i} \ll \lambda_{i,j}$ the Legget mode is in the continuum above $2\Delta$ and is therefore overdamped. Figure~\ref{fig:Leggett} shows the transition between the two coupling regimes and demonstrates the damping effect. In addition, it could be shown that the energy saturates at the maximal gap energy.

\subsection{Spin excitations}
\label{sec:spin}
For spin excitations the Elliot-Fleury-Loudon Hamiltonian \cite{Fleury:1968} is the simplest nonresonant interaction operator which desribes scattering in a Heisenberg model with nearest-neighbour exchange coupling $J$,
\begin{equation}
  \label{eq:FL}
  \hat{H}_{\rm EFL} = J\sum_{<i,\hat{{\bf \delta}}>}(\hat{\bf \delta}\cdot\hat{\bf e}_I)(\hat{\bf \delta}\cdot\hat{\bf e}_S)({\bf S}_{{\bf r}(i)}\cdot{\bf S}_{{{\bf r}(i)} + \hat{\delta}a}).
\end{equation}
${\bf S}_{{\bf r}(i)}$ is a spin at site ${{\bf r}(i)}$, $a$ is the distance between the sites and $\hat{\bf \delta}$ is a unit vector pointing towards one of the nearest neighbors. $<i,\hat{\bf \delta}>$ is a restricted sum to avoid double counting. The spectral shape can either be determined in terms of spin-wave theory \cite{Fleury:1968}, by numerical \cite{Chen:2011b} or field theoretical methods \cite{Weidinger:2015}.

Screening may become relevant in metallic systems such as the FeBCs in which the spins forming the SDW are presumably itinerant as opposed to cuprates which are Mott insulators at low doping. However, there is no analytic treatment yet dealing with the problem of light scattering from spin polarized conduction electrons beyond the nearly antiferromagnetic Fermi liquid \cite{Kampf:1999}. Using Quantum Monte Carlo methods, doped cuprates were studied \cite{Moritz:2011,Jia:2014}. Considerations along these lines may surface when analyzing the differences between the pnictides and chalchogenides where the magnetism is believed to be predominantly itinerant and localized, respectively, at least for some orbitals \cite{Yin:2011,Georges:2013,Si:2016,WangQS:2016,Skornyakov:2017}. However, this rather general problem cannot be solved here although the controversial discussion on spin excitations in the FeBCs requires additional input. 

\subsection{Fluctuations}
\label{sec:fluctuations}

In the case of fluctuations, there are three possibilities to deal with the $q=0$ selection rule. (i) A fluctuation can have zero momentum. This case applies if all unit cells have the same excitation pattern as for the case of ferro-orbital fluctuations \cite{Thorsmolle:2016}. (ii) A fluctuation with finite critical wave vector such as ${\bf q}_c = (\pi,\pi)$ or ${\bf q}_c = (\pi,0)$ for incipient N\'eel type or stripe-like antiferromagnetic order, respectively, can exchange momentum of opposite sign with another excitation. This case may work in the presence of moderately strong impurity concentrations \cite{Gallais:2016a}. The momentum transferred from the light is certainly insufficient for large critical momenta (see section \ref{sec:LS-metals}). (iii) Two fluctuations with opposite momenta are exchanged. It is exactly what happens in the case of two-magnon excitations in a system with long-ranged order \cite{Fleury:1968,Sulewski:1991}. The same type of scattering may also occur in partially ordered systems \cite{Venturini:2000} or in the presence of critical fluctuations \cite{Caprara:2005}. This type of diagrams were first studied by Aslamazov and Larkin in the context of paraconductivity above the superconduction transition \cite{Aslamasov:1968}.

For the FeBCs there are various theoretical studies of fluctuations \cite{Khodas:2015,Karahasanovic:2015}. Although spin, orbital, and charge degrees of freedom are not independent, the question as to the leading instability remains relevant and crucial for the understanding of the FeBCs \cite{Thorsmolle:2016,Karahasanovic:2015,Gallais:2016a,Khodas:2015,Fernandes:2012,Kretzschmar:2016}.

\subsection{Selection rules}
\label{sec:selection rules}

In the inelastic light scattering experiment, all selection rules must be compatible with the direct product of two dipole transitions in the relevant crystal structure \cite{Hayes:2005}. We consider only those for spin and charge excitations since they are well-known for phonons.

\begin{figure}
  \centering
  \includegraphics [width=8.5cm]{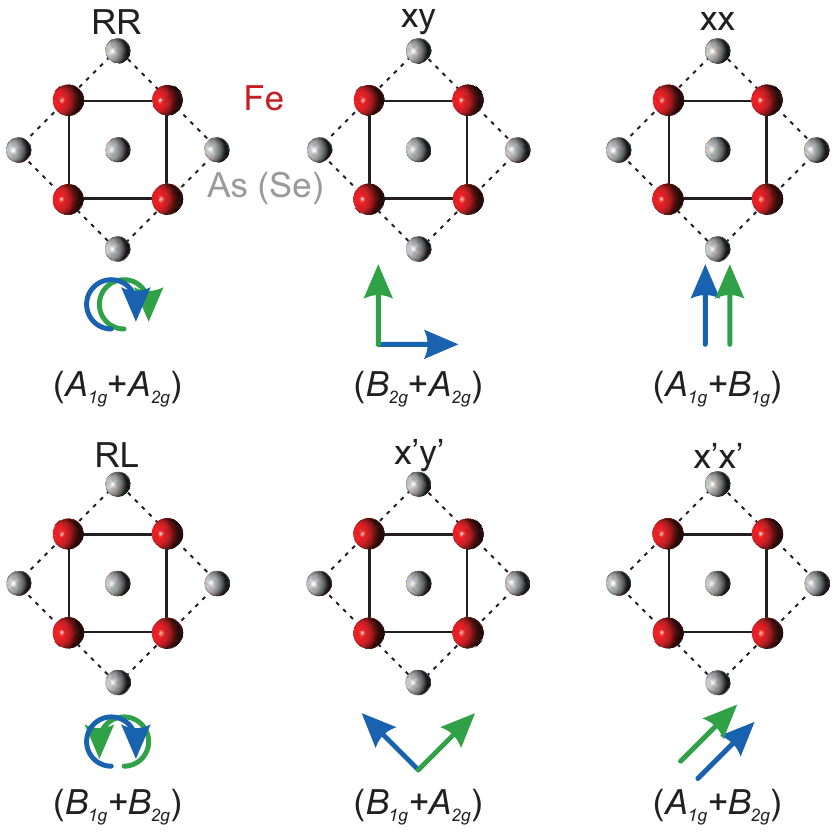}
  \caption[]{Scattering geometry and symmetries for the $ab$ plane of an FeBCs. Incoming and scattered photons are indicated as blue and green arrows. In backscattering configuration, the arrows corresponding to $R$ and $L$ polarization of the scattered light should be interchanged. The symmetries refer to the 1\,Fe unit cell (full line) which is relevant (and frequently used) for electronic and spin excitations. Phonons have the right symmetry in the crystallographic or 2\,Fe unit cell (broken line) where \Blg and \BZg are interchanged with respect to the 1\,Fe cell.}
  \label{fig:srules}
\end{figure}

\subsubsection{Particle-hole and gap excitations}

\begin{figure}
  \centering
  \includegraphics [width=8.5cm]{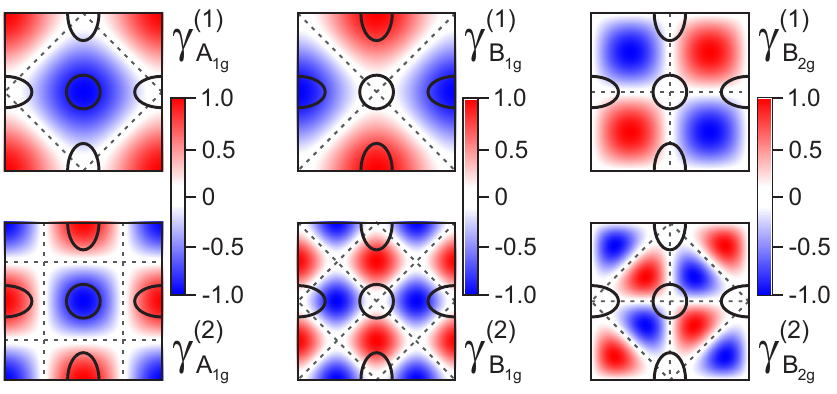}
  \caption[]{Symmetry properties and momentum dependences of the Raman vertices $\gamma_{\mu}$. Shown are the first and second order vertices for polarization orientations transforming as $\mu = A_{1g}$, $B_{1g}$ and $B_{2g}$ for the $D_{4h}$ space group. The zeroth order $A_{1g}$ vertex is just a constant and is entirely screened. Higher order $A_{1g}$ vertices are only partially screened. The solid lines represent an idealized two-band Fermi surface of FeSCs in the 1\,Fe unit cell. For the symmetry properties of the band structure in the FeBCs the screening in $A_{1g}$ symmetry is almost negligible \cite{Boyd:2009}.
  }
  \label{fig:vertices}
\end{figure}

In the simplest approximation the vertices can be expanded into crystal harmonics of the respective point group of the crystal \cite{Allen:1976}. Figures~\ref{fig:srules} and \ref{fig:vertices} summarize the main scattering geometries and sensitivity projections, resepectively, of a $D_{4h}$ point symmetry, relevant for FeBCs and cuprates. From linear combinations of the spectra measured in these geometries, pure symmetries $\mu$ can be derived.

In the first order approximation selection rules can be derived on the basis of symmetry alone since $\gamma_{\mu}(\bf{k},\bf{q}\rightarrow 0)$ can be expanded into the set of basis functions $\Phi_{\mu}(\bf{k})$ \cite{Einzel:1996}. Figure~\ref{fig:vertices} shows schematic representations of the first and second order basis functions (crystal harmonics) of each symmetry to which the Raman vertices $\gamma_{\mu}$ are proportional.
For particle-hole excitations, symmetry-resolved Raman vertices can be derived from the band structure within the effective mass approximation,

\begin{equation}\label{Eq:A1g}
\gamma_{A_{1g}}({\bf k},{\bf q}\rightarrow 0)\propto\frac{1}{2}\big(\frac{\partial^2 E_{{\bf k}}}{\partial k_xk_x}+\frac{\partial^2 E_{{\bf k}}}{\partial k_yk_y}\big)
\end{equation}

\begin{equation}\label{Eq:B1g}
\gamma_{B_{1g}}({\bf k},{\bf q}\rightarrow 0)\propto\frac{1}{2}\big(\frac{\partial^2 E_{{\bf k}}}{\partial k_xk_x}-\frac{\partial^2 E_{{\bf k}}}{\partial k_yk_y}\big)
\end{equation}

\begin{equation}\label{Eq:B2g}
\gamma_{B_{2g}}({\bf k},{\bf q}\rightarrow 0)\propto\big(\frac{\partial^2 E_{{\bf k}}}{\partial k_xk_y}\big),
\end{equation}
where, as in the case of FeBCs, ${\bf k}$ effectively takes into account all contributions from different bands. This type of calculation was conducted for \BFCA and \BKFA in Refs. \cite{Mazin:2010a,Bohm:2014} yielding $\gamma_{A_{1g}}^{(2)}$ rather than $\gamma_{A_{1g}}^{(1)}$  to be relevant for presumably most of the FeBCs.

If, for the sake of simplicity, we omit the contribution from any additional excitations (e.g. phonons) as well as interactions, the first-order diagram on the right hand side of Fig.~\ref{fig:feynman}\,(a), which describes particle-hole excitations, yields the full Raman response in the normal state. Although the propagators in the SC state are different from those above $T_c$, the first order diagram is still sufficient for describing the response below $T_c$. Their evaluation yields a Raman response which depends quadratically on $\gamma_{\mu}({\bf k})$. As can be seen from Fig.~\ref{fig:vertices}, in the \Blg symmetry ($\gamma_{B{1g}}^{(1)}$ ) one may expect to probe electron pockets, whereas both electron and hole pockets may be probed in \Alg symmetry ($\gamma_{A{1g}}^{(2)}$). The dependence of the Raman response on $\gamma^2_{\mu}({\bf k})$ directly explains the selectivity in $k$-space.

\subsubsection{Bardasis Schrieffer modes}
The selection rules for BS modes depend on the symmetry of the sub-leading interaction. The first proposals considered a ground state being driven by spin fluctuations thus having an $s$-wave gap with opposite sign on the electron and hole bands. Then, if the sub-leading channel originates from orbital fluctuations, predominantly along the $(\pi,0)$ direction \cite{Kontani:2010}, the BS mode is expected in the $A_{1g}$ channel \cite{Chubukov:2009}. Alternatively, the next strongest pairing interactions can have the same origin as the ground state resulting from spin fluctuations between the electron bands along $(\pi,\pi)$ thus entailing a BS mode in $B_{1g}$ symmetry.

\subsubsection{Leggett modes}
The symmetry selection rules for the Leggett modes depend on the interactions included and thus on the band structure and the coupling. If two concentric bands or interactions with wave-vector $(\pi,0)$ are included the Leggett mode appears in the fully symmetric $A_{1g}$ channel \cite{Cea:2016}. In the case when $(\pi,\pi)$ interactions have to be considered, in particular if the central hole band is missing, the Leggett modes appear in $B_{1g}$ symmetry \cite{Huang:2018}. In the case of fluctuations between the $d_{xz}$ and $d_{yz}$ orbitals Leggett modes may also appear in $B_{1g}$ symmetry so long as they are not overdamped \cite{Burnell:2010}.

\subsubsection{Fluctuations}

Fluctuation contributions to the Raman response may arise from one and/or two fluctuations [see Fig.~\ref{fig:feynman}\,(b)]. As can be seen from the diagrams, the first order term is non-vanishing only if the symmetry of the fluctuation is that of the Raman vertex. On the other hand, second order diagrams include electronic loops $\Lambda_{\mu}^0(\bf q)$ which depend linearly on $\gamma$ and quadratically on $f$. Consequently, for a given critical vector ${\bf q}_c$ and a set of FS hot-spots ${\bf k}_0$ which are connected by ${\bf q}_c$, the selection rules for the second order term read $\Lambda_{\mu}^0({\bf q}_c)\propto \sum_{{\bf k}_0}\gamma_{\mu}({\bf q}_c)$ \cite{Caprara:2005}. It states that a finite response in symmetry channel $\mu$ is expected only if ${\bf q}_c$ connects hot-spots in which $\gamma_{\mu}$ does not change sign. As opposed to the first order term, $\gamma_{\mu}$ does not necessarily reflect the symmetry of the fluctuation for second order processes.

\section{Instabilities beyond superconductivity}

The phase diagram (Fig.~\ref{fig:structure}) shows that the FeBCs have a host of instabilities beyond superconductivity.  Their interrelation amongst each other and with superconductivity is a main focus of research. For certain ranges of doping and/or applied pressure all FeBCs exhibit long range magnetism while remaining metallic in stark contrast to the cuprates. Some of the FeBCs like FeSe display only short-ranged magnetic order at finite temperature \cite{WangQS:2016,Baek:2014}. Not surprisingly, the nature of the magnetism is still under discussion.  It was shown recently that the degree of localization of electrons and thus  spins may depend on the orbital and varies between the materials \cite{Yin:2011}.

Above the magnetic ordering temperatures there are various structural transitions which typically go along with electronic anisotropies such as substantial differences in the resistivities measured along inequi\-valent directions \cite{Chu:2010} or orbital order, specifically of the $d_{xz}$ and $d_{yz}$ orbitals \cite{Yi:2011}. Yet, how can the origin of these differences be pinned down?

\subsection{Excitations from localized spins}
\label{sec:local-spin}

While itinerant and localized magnetism cannot easily be distinguished by neutron scattering, Raman scattering offers clear criteria \cite{Baum:2019} which are outlined in sections \ref{sec:weak} and \ref{sec:spin}. The main arguments are the temperature, doping and symmetry dependences and line shapes.

Early experiments on Fe chalcogenides and pnictides reported excitations in the range 2\,000 to 4\,000\,cm$^{-1}$ in all symmetries. The results were interpreted in terms of local spins \cite{Sugai:2012,Sugai:2011}, and the energies are in fact compatible with the exchange coupling $J\approx 120$\,meV found by neutron scattering or density functional theory (DFT) \cite{Glasbrenner:2015}. With the improvement of the sample quality, the peaks in this range faded away and may be traced back to luminescence with high probability \cite{Y.X.Yang:2014}.

Recent experiments in FeSe support scattering from localized spins. The $B_{1g}$ spectra at high (300\,K) and low (20\,K) temperature are dominated by the broad peak centred at approximately 500\,cm$^{-1}$, whereas near the $T_S$ an additional sharp peak appears in the range 100-200\,cm$^{-1}$ \cite{Massat:2016,Zhang:2017s,Baum:2019} which we disregard here but discuss in detail below in the context of fluctuations. The main peak in $B_{1g}$ symmetry at 500\,cm$^{-1}$ depends continuously on temperature \cite{Baum:2019} and survives low doping with sulfur \cite{Zhang:2017s}. For FeS the excitation is completely gone. In contrast to Massat \textit{et al.} \cite{Massat:2016} and Zhang \textit{et al.} \cite{Zhang:2017s}, Baum and coworkers \cite{Baum:2019} assign this peak to a two-magnon excitation of nearly frustrated spins.

There are experimental and theoretical arguments for this assessment. Very generally, antiferromagnetically ordered local spins give rise to two-magnon excitations close to $3J$ \cite{Fleury:1968} whereas SDW order in a metallic system leads to instantaneously appearing coherence effects close to the gap energy in the electronic excitation spectrum similar to those in a superconductor (see section \ref{sec:weak}). In FeSe, the pronounced excitation in \Blg symmetry on top of the particle-hole continuum builds up gradually \cite{Massat:2016,Baum:2019}. Thus the temperature dependence of the scattering intensity is compatible with that of a quasi two-dimensional N\'eel antiferromagnet, e.g., YBa$_2$Cu$_3$O$_6$ \cite{Knoll:1990}. However, the energy of the \Blg peak is too small, by approximately a factor of five, suggesting frustrated magnetism as studied theoretically already in pnictides and chalcogenides \cite{Chen:2011b}, expected from LDA results \cite{Glasbrenner:2015}, and explored recently for FeSe using exact diagonalization \cite{Ruiz:2019}.

On a square lattice, frustration can occur in the presence of interactions beyond nearest-neighbour coupling. In this case, a variety of ordering patterns or wave vectors may be realized. A possible model is the $J_1-J_2-J_3-K$ Heisenberg model \cite{Glasbrenner:2015} where nearest, next-nearest and next-next-nearest neighbour exchange couplings are taken into account. $K$ is the coefficient of the bilinear interaction proportional to $({\bf S}_i\cdot{\bf S}_j)^2$ which, depending on the sign (de)stabilizes long-range order. For $J_2\approx 0.5J_1$ N\'eel $(\pi,\pi)$ and stripe $(\pi,0)$ order occur with similar probability for $J_3\to 0$ and are separated by a $(\pi/2,0)$ phase for $J_3>0$ \cite{Glasbrenner:2015}. In all cases, little energy is required for flipping a spin, and the maximum of the two-magnon excitation moves to zero energy in the classical limit ($S\to\infty$) and to approximately $0.5J_1$ for $S=1/2$ as opposed to $2.84J_1$ when $J_1$ dominates \cite{Weidinger:2015}. The near degeneracy of the N\'eel and the stripe state going along with short range order in FeSe was not only predicted by LDA studies \cite{Glasbrenner:2015} but also observed recently by neutron scattering \cite{WangQS:2016}.

For the low energy and the temperature dependence of the \Blg excitation in FeSe, the response was studied also theoretically by diagonalizing a $4\times4$ cluster carrying spin 1 using $J_2=0.528J_1$ and $J_3=0$ \cite{Baum:2019,Ruiz:2019}. The observed agreement between theory and experiment is semi-quantitative for all symmetries. The conclusions are in agreement with those of the neutron scattering experiments  \cite{WangQS:2016} and support the existence of local spins. How are local spins compatible with metallic transport in FeSe and why is FeSe different from the pnictides? Before discussing this question in section~\ref{sec:origin} the SDW systems and fluctuations will be reviewed.

\subsection{Spin density wave order}
\label{sec:SDW}

\begin{figure}
  \centering
  \includegraphics [width=8.5cm]{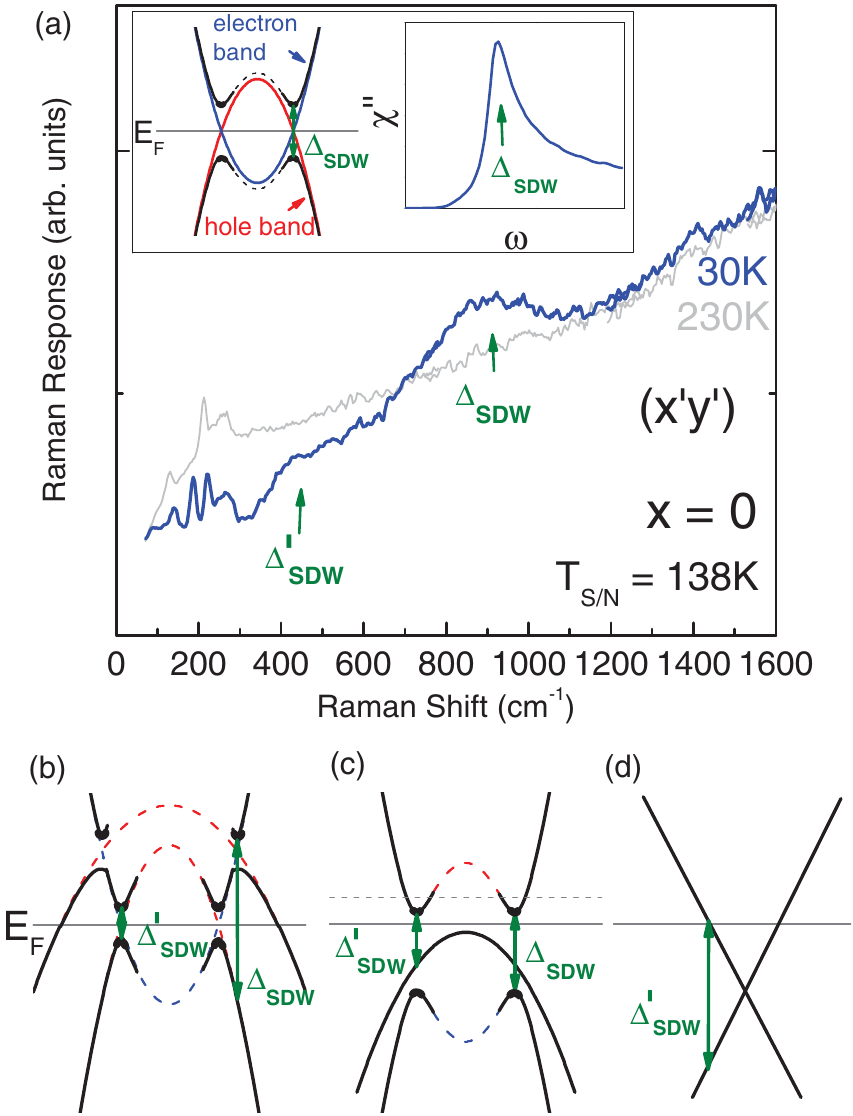}
  \caption[]{(Color online) Effect of SDW formation on the Raman response of BFA. (a) High- and low-temperature Raman response in $x^\prime y^\prime$ configuration. The spectra show two spectral features at and $\Delta^\prime_{\rm SDW}$. (b) If one electron band and two hole bands anti-cross $\Delta_{\rm SDW}$ is much larger than $\Delta^\prime_{\rm SDW}$. (c) If one of the hole bands does not reach the Fermi level the two gaps are expected to have similar magnitude. For phase-space reasons $\Delta^\prime_{\rm SDW}$ should have much stronger spectral weight than $\Delta_{\rm SDW}$ in case (b) than in case (c) and \textit{vice versa}. (d) In the presence of a Dirac cone (non-interacting bands) $\Delta^\prime_{\rm SDW}$ should be characterized by very small spectral weight and $\Delta_{\rm SDW}$ is not expected to be observable. From \cite{Chauviere:2011} with permission.
  }
  \label{fig:Chauviere:2011}
\end{figure}

Signatures of the SDW in the Raman spectra of BFA were first discussed by Chauvi\`{e}re \textit{et al.}~\cite{Chauviere:2011}. Figure~\ref{fig:Chauviere:2011} compares the low and high temperature $B_{1g}$ Raman response. Spectral weight is redistributed from energies below to above the SDW gap upon entering the SDW state. The Raman spectra exhibit two distinct features: A peak at about 900\,cm$^{-1}$ appearing only in B$_{1g}$ symmetry and a step-like increase at about 400\,cm$^{-1}$ in all channels \cite{Chauviere:2011}.

The $B_{1g}$ selection rules for inter-orbital transitions would be compatible with a transition $d_{x^2-y^2}\leftrightarrow d_{z^2}$ to correspond to the peak at 900\,cm$^{-1}$. Yet, this straightforward explanation is in conflict with band-structure calculations which find the dominant contribution to the states at the Fermi surface to originate from $d_{xz}$, $d_{yz}$, and $d_{xy}$ orbitals. Thus intensity can only be redistributed among these orbitals. Consequently, a band-folding picture was suggested with two types of electronic transitions in the SDW state: (i) a high-energy transition between electron and hole bands anti-crossing after back-folding and (ii) a low-energy transition involving either interacting or noninteracting bands. Upon Co substitution, the peak observed in $B_{1g}$ symmetry disappears due to the increase of the Fermi energy and the related filling of the unoccupied states at the anti-crossing points \cite{Chauviere:2011}.
\begin{figure*}
  \centering
  \includegraphics [width=12cm]{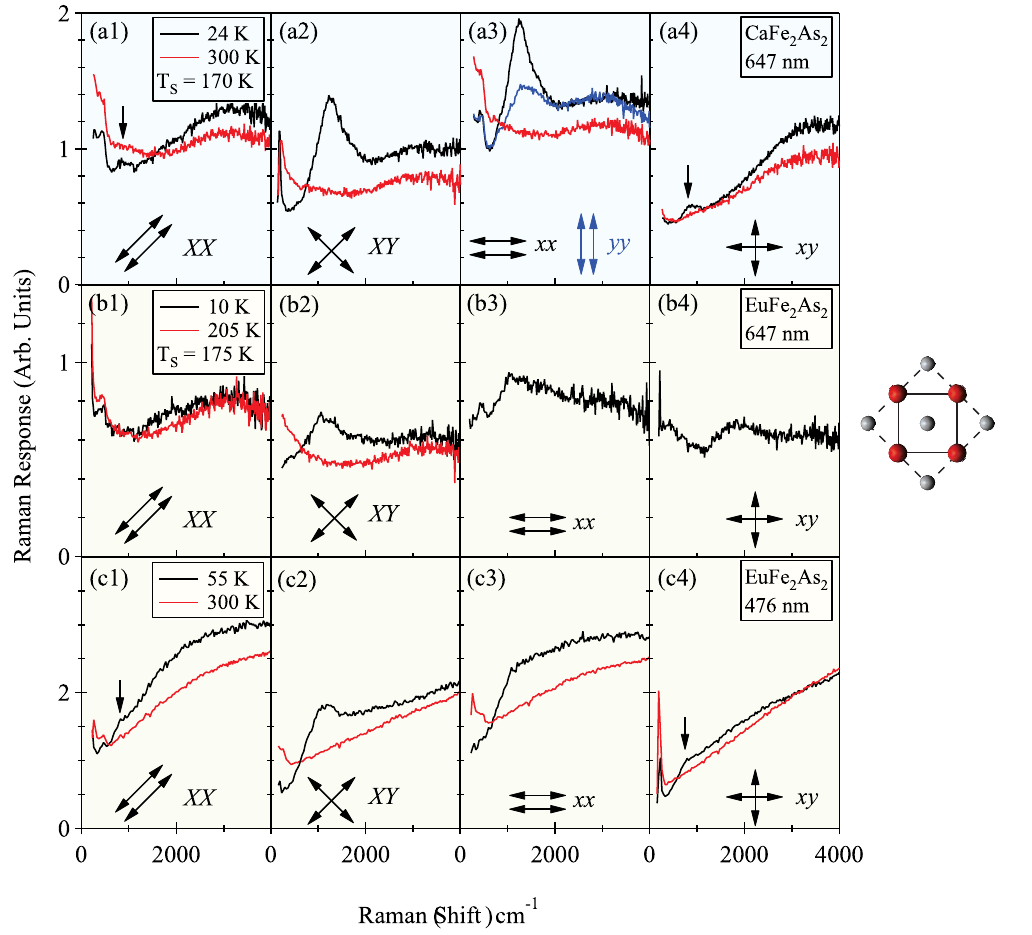}
  \caption[]{(Color online) Effect of SDW order in the Raman spectra of CaFe$_2$As$_2$ and EuFe$_2$As$_2$. The unit cell on the right to which the polarizations in the panels refer is added to the original figure for clarity. (a1)-(a4) Spectra CaFe$_2$As$_2$ measured with 647\,nm laser excitation in the normal state (300\,K, red) and in the de-twinned SDW phase (24\,K, black/blue). (b1)-(b4) Raman response of EuFe$_2$As$_2$ in the normal state (205\,K, red) and the twinned SDW state (10\,K, black). (c1)-(c4) Same as (b1)-(b4) with the 476\,nm laser excitation. Note the notation $X\equiv x^\prime$ and $Y\equiv y^\prime$. From \cite{W.L.Zhang:2016} with permission.
  }
  \label{fig:Blumberg}
\end{figure*}

A much more pronounced redistribution of spectral weight was reported for Sr122 \cite{Y.X.Yang:2014} with a suppression at low energies and three distinct peaks at 820\,cm$^{-1}$, 1140\,cm$^{-1}$ and 1420\,cm$^{-1}$ appearing in $B_{1g}$ symmetry. The peak at 820\,cm$^{-1}$ is also present in $B_{2g}$ symmetry. In $A_{1g}$ symmetry no peaks can be resolved. On the basis of the symmetry selection rules the authors argue that the peaks at 1140\,cm$^{-1}$ and 1420\,cm$^{-1}$ originate from anti-crossing bands in the presence of imperfect nesting rendering the $X$ and $Y$ points inequivalent. The peak at 1420\,cm$^{-1}$ appears only well below $T_N$ and may be understood in terms of a temperature dependent Fermi surface topology and the disappearance of a hole-like Fermi surface pocket very close to the chemical potential in the reconstructed SDW electronic structure. The peak at 820\,cm$^{-1}$ is assigned to an optical transition between folded bands away from the $\Gamma - X$ and $\Gamma - Y$ directions which are probed in $xy$ configuration.

\begin{figure*}
  \centering
  \includegraphics [width=12cm]{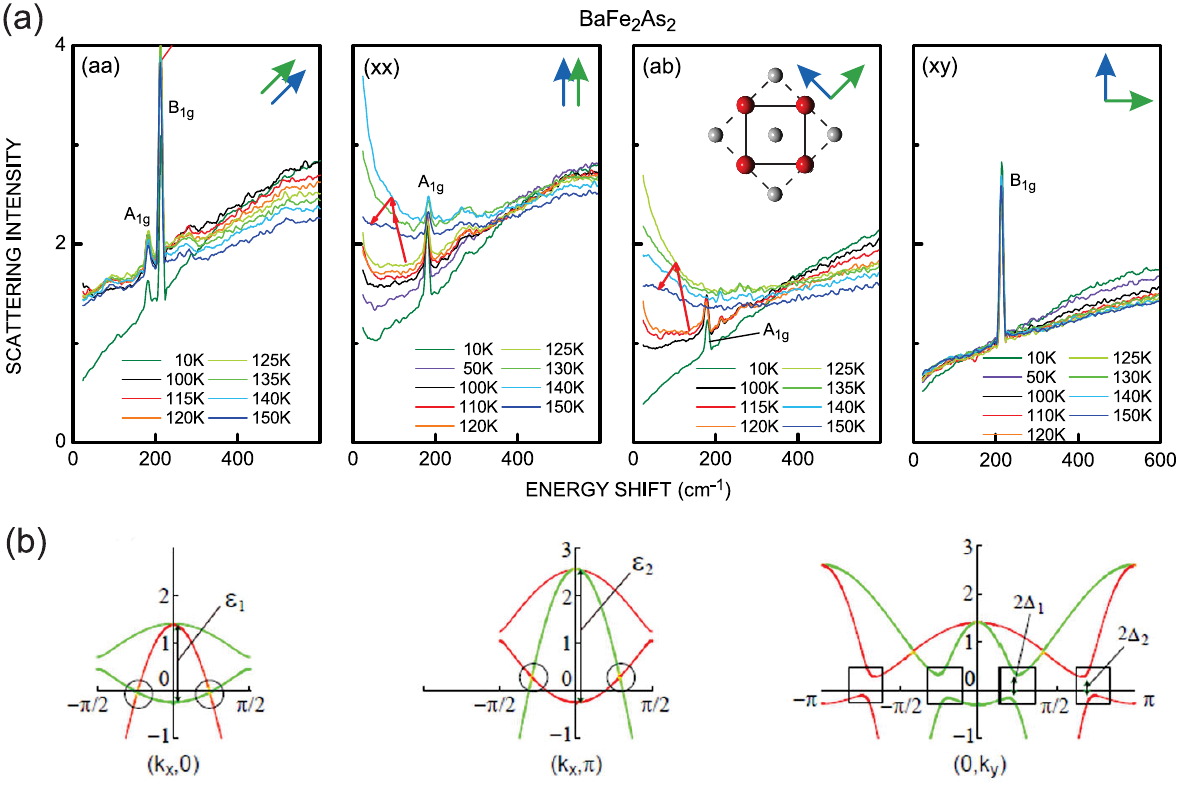}
  \caption[]{(Color online) (a) Low-energy Raman spectra of BaFe$_2$As$_2$ at temperatures as indicated. The pictograms displaying the unit cell and the polarizations are added to the original figure for clarity. The two central panels $(xx)$ and $(ab~[\equiv x^\prime y^\prime]$ show contributions from fluctuations at low energy and the redistribution of spectral weight from below to above 350\,cm$^{-1}$ upon cooling through $T_{\rm SDW} \approx 135$\,K as indicated by red arrows. (b) Electron dispersions in $(k_x,0)$, $(k_x,\pi)$, and $(0,k_y)$ direction. The circles denote the Dirac nodes, and the squares denote the ``anti-nodes''. From \cite{Sugai:2012,Sugai:2013} with permission.
  }
  \label{fig:Sugai}
\end{figure*}

Recently  Zhang \textit{et al.} \cite{W.L.Zhang:2016} reported  spectral weight redistribution in twinned Eu122 and detwinned mono-domain Ca122 single-crystals in the SDW phase, as shown in  Fig.~\ref{fig:Blumberg}. In the $x'y' ~[\equiv XY]$ configuration, spectral weight is transfered from low energy to above 800\,cm$^{-1}$ with the development of a peak at 1220\,cm$^{-1}$ and 1060\,cm$^{-1}$ for Ca122 and Eu122, respectively [see panels (a2) and (b2) in Fig.~\ref{fig:Blumberg}]. For de-twinned Ca122 a large intensity anisotropy of the 1220\,cm$^{-1}$ peak is observed in $xx$ and $yy$ scattering configurations [see panel (a3) in Fig.~\ref{fig:Blumberg}]. In addition, a weak spectral feature is observed  at 830\,cm$^{-1}$ in $x'x' ~[\equiv XX]$ and $xy$ scattering configurations as indicated by arrows. The authors derive the selection rules for inter- and intra-orbital transitions on the basis of the $D_{2h}$ group for the high symmetry points in the Brillouin zone. The symmetry analysis augmented by an orbital-resolved DFT+DMFT study \cite{W.L.Zhang:2016} suggests that the peak at 1200\,cm$^{-1}$ in the $A_g$ symmetry channel originates from an intra-orbital transition at the $Z$ point induced by the SDW band folding, whereas the peak at 830\,cm$^{-1}$ in the $B_{1g}$ channel arises from the $d_{xz} \leftrightarrow d_{yz}$ transition at the $\Gamma$ point.

The analysis of the Raman spectra in the SDW state shows that the typical gaps induced by ordering of the magnetic moments of itinerant electrons are in the range of 100-150\,meV  or 8\,$k_{\rm B}T_{\rm SDW}$. This ratio is in the same range as that of the superconducting gap, and one cannot conclude that $T_{\rm SDW}$ is suppressed by fluctuations, as, for instance in the tritellurides \cite{Eiter:2013}. Rather, the large value may indicate strong coupling. Whether or not the $a-b$ anisotropy of the gap energy found in Ca122 \cite{W.L.Zhang:2016} is a generic feature of all pnictides cannot be decided on the basis of the material at hand. A strong electronic anisotropy was also found in the energy range of 2-3\,eV in Ba122 below \TSDW when studying resonance effects of the $A_g$ arsenic phonon \cite{Baum:2018a}. Thus, anisotropies in the electronic structure were identified by Raman scttering at low and high energies in the magnetically ordered phase.

\subsection{Fluctuations above the ordering transitions}
\label{sec:fluct-exp}

\begin{figure*}
  \hspace*{2cm}
  \includegraphics [width=12cm]{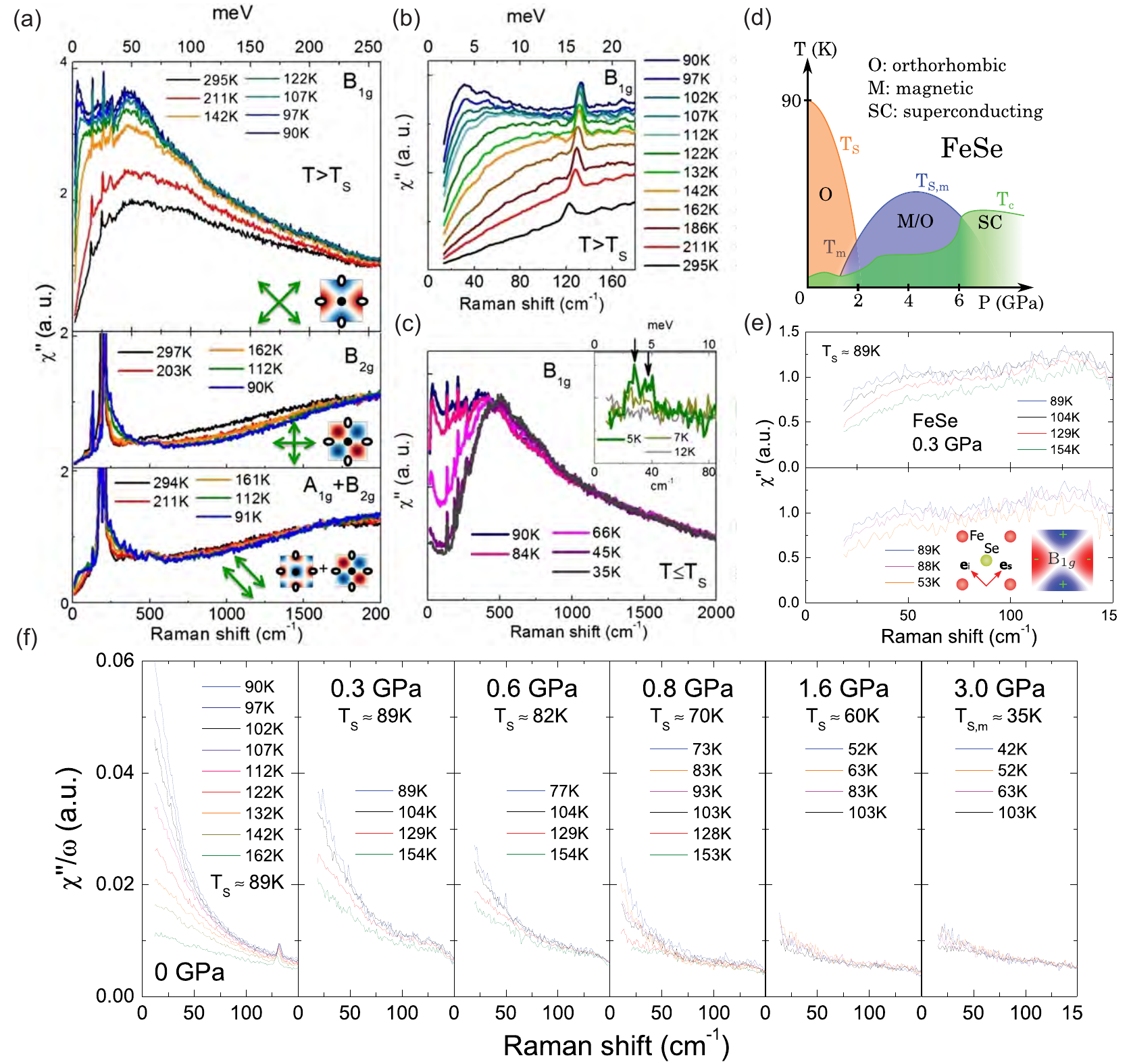}
  \caption[]{(Color online) Light scattering in FeSe. (a)-(c) Symmetry-dependent Raman spectra of FeSe above $T_s$=87 K using photons at 2.33\,eV. The sharp peaks superimposed on the electronic continuum are due to Raman active optical phonons. The insets display the Raman form factors in different symmetries (blue and red colors indicate positive and negative amplitudes, respectively), and the polarization configurations used to select them. (b) Temperature dependence of the low energy B$_{1g}$ spectra above $T_s$. (c) Evolution of the B$_{1g}$ spectra across $T_s$. The Inset shows the spectra across the superconducting transition at $T_c$=8.5 K. The arrows indicate the peaks associated with 2$\Delta$. From \cite{Massat:2016}. (e) Pressure-temperature hase diagram of FeSe.
  }
  \label{fig:Massat}
\end{figure*}

There are various types of instabilities in the FeSCs which can drive the phase transitions. While some groups consider magnetic ordering the dominant interaction \cite{Mazin:2008,Fernandes:2012a}, Kontani \textit{et al.} see orbital ordering in the driver's seat \cite{Kontani:2011}. Since strong fluctuations precede the structural transitions in many compounds, experimental access to the fluctuations is highly desirable. Raman spectroscopy is among the  handful techniques available but does not probe the fluctuations independently. Rather, since the FeBCs are metals the contribution is always superimposed on the particle-hole continuum which needs to be taken into account for quantitative analysis.

A contribution from fluctuations to the Raman spectra of FeBCs were first reported by Choi \textit{et al.} \cite{Choi:2010}. They observed a pronounced build up of the low energy Raman response in Sr122  upon cooling towards \Ts and suggested a magnetic nature of the fluctuations. The similar build up of the low-energy signal in the $B_{1g}$ channel was also observed for Ba122 by Sugai \textit{et al.} \cite{Sugai:2012,Sugai:2013} as presented in Figure~\ref{fig:Sugai}. The authors attributed the low-energy $x^\prime x^\prime ~[\equiv aa]$ and $xy$ spectra to excitations near Dirac nodes where the bands intersect without interacting [circles in Fig.~\ref{fig:Sugai}\,(b)] and the $xx$ and $x^\prime y^\prime ~[\equiv ab]$ spectra to the ``anti-nodal'' excitations where the back-folded bands interact [squares in Fig.~\ref{fig:Sugai}\,(b)]. The increase of the low-energy scattering intensity was interpreted in terms of critical fluctuations related to the opening of the anti-nodal gap.

Several detailed studies of fluctuations were presented more recently in Refs. \cite{Gallais:2013,W.L.Zhang:2014,Kretzschmar:2016,Thorsmolle:2016,Gallais:2016a,Massat:2016,%
Gnezdilov:2013,WuSF:2016a}. As discussed in section~\ref{sec:fluctuations}, one and/or two fluctuations may contribute to the Raman spectra via distinct scattering processes. Gallais and Paul \cite{Gallais:2016a} and Thorsm{\o}lle \textit{et al.} \cite{Thorsmolle:2016} point out that the first order term has a non-zero contribution only in the presence of momentum scattering processes (or, equivalently, finite $v_Fq$). On the other hand, the second order AL diagrams may include  contributions having $q\gg 0$. 

Gallais and coworkers suggested to use the static limit of the real part of the response, $\chi^\prime_{B1g}(\Omega= 0, T)$, for analyzing the Raman data and comparing them to the results of other quasi-static methods such as NMR or elasticity. $\chi^\prime_{B1g}(\Omega= 0, T)$  was directly extracted from $\chi^{\prime\prime}_{B1g}(\Omega, T)$ \textit{via} Kramers-Kroning transformation. This procedure requires an extrapolation of $\chi^{\prime\prime}_{B1g}(\Omega, T)$ to zero energy and the selection of a high-energy cutoff since $\chi^{\prime\prime}_{B1g}(\Omega\to \infty, T)\approx c$ where $c$ is a constant. The method was applied to the analysis of the data of BFCA \cite{Gallais:2013,Gallais:2016a} (see next paragraph) and subsequently of Eu/Sr122 \cite{W.L.Zhang:2014}, BKFA \cite{WuSF:2016a}, and Na111 \cite{Thorsmolle:2016}. Thorsm{\o}lle \textit{et al.} demonstrated the scaling of the static susceptibility obtained from Raman scattering and NMR data for Na111. Consequently, low energy quasielastic scattering was related to $d^{\pm}$ quadrupolar nematic fluctuations which became critical on approaching a Pomeranchuk instability with a deformation of the Fermi surface \cite{Yamase:2011,Yamase:2013}.

The low-energy Raman response in differently doped BFCA was extensively studied by Gallais \textit{et al.}  \cite{Gallais:2013,Gallais:2016a}. The authors argue that the nematic susceptibility is observable in the $B_{1g}$ channel and originates from charge fluctuations for symmetry reasons since inelastic light scattering couples preferably to the charge. They observe a strong enhancement of the $B_{1g}$ response $\chi^{\prime\prime}_{B1g}(\Omega\approx k_B T, T)$ upon cooling towards \Ts and a collapse thereof in the orthorhombic/SDW state, whereas the $B_{2g}$ response is essentially temperature independent. First, $\chi^{\prime}_{B1g}(\Omega= 0, T)$  was determined. Second, it is shown that this quantity is equal to the quasi-static nematic susceptibility and, in fact, compares well to the Young modulus $c_{66}(T)$ derived from thermal expansion \cite{Gallais:2016a,Yoshizawa:2012}. Interestingly, neither $c_{66}(T)$ nor $\chi^{\prime}_{B1g}(\Omega= 0, T)$ diverge at $T_{\rm s}$ due to the coupling of the electronic/spin degrees of freedom to the lattice \cite{Gallais:2013,Gallais:2016a,Kontani:2011}. As a consequence, the divergence of $\chi^{\prime}_{B1g}(\Omega= 0, T)$ derived from a Curie-Weiss fit occurs always below the structural transition, $T_0<\Ts$.

\begin{figure*}
  \centering
  \includegraphics [width=12cm]{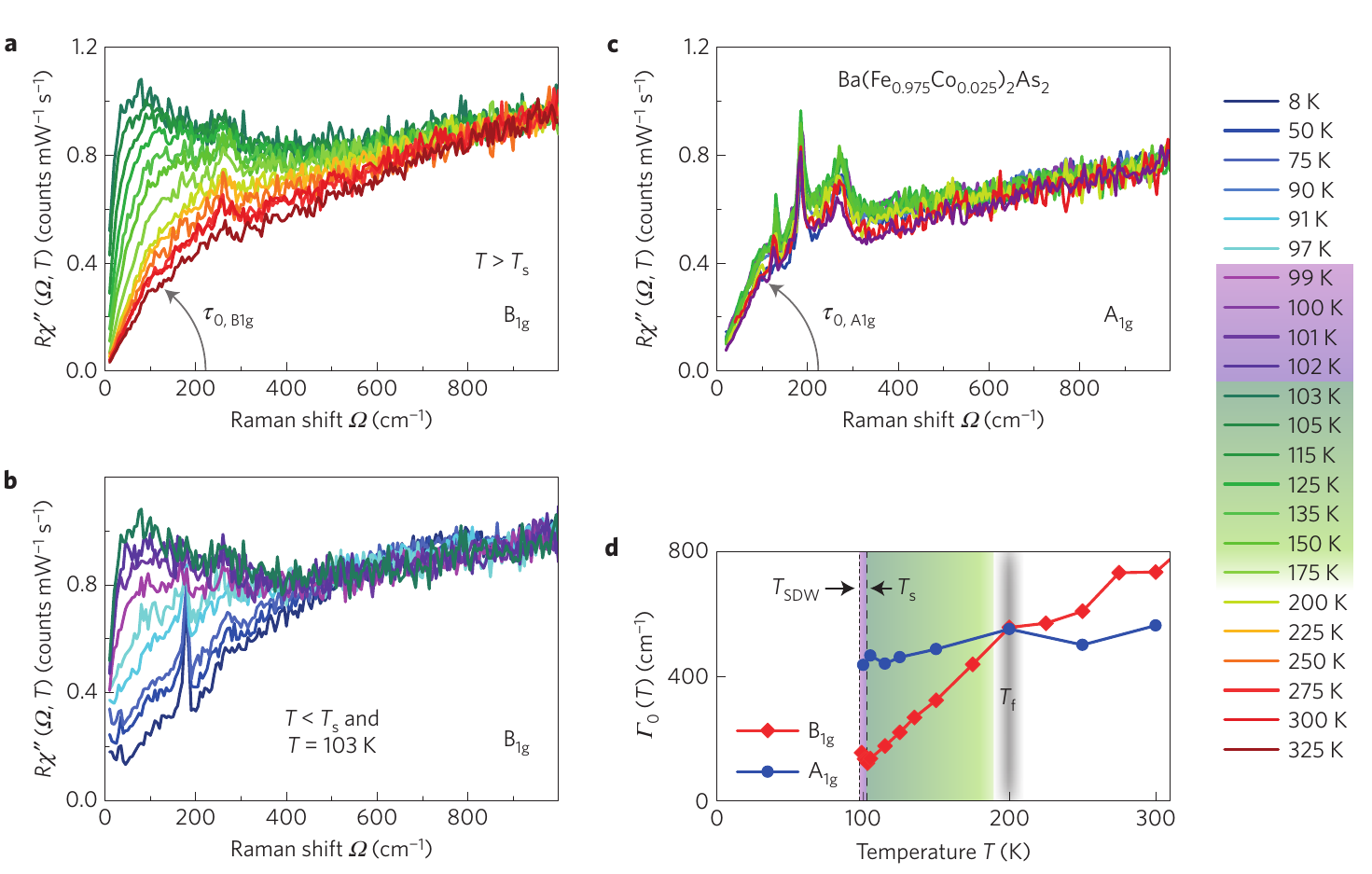}
  \caption[]{(Color online) Polarization-resolved Raman results for ${\rm Ba(Fe_{0.975}Co_{0.025})_2As_2}$. (a)–(c) Response $R\chi''(\Omega,T)$ (raw data after division by the Bose–Einstein factor) at temperatures as indicated. (a) $B_{1g}$ spectra above and (b) below \Ts and (c) $A_{1g}$ symmetry. The initial slopes defined in (a) and (c) as grey arrows are proportional to the static two-particle lifetime $\tau_{0,\mu}$ in symmetry $\mu=A_{1g}$, $B_{1g}$. (d) Raman relaxation rates $\Gamma_{0,\mu}(T)=\hbar/\tau_{0,\mu}(T)$, in $A_{1g}$ (blue circles) and $B_{1g}$ (red diamonds) symmetry as a function of temperature. The $A_{1g}$ and $B_{1g}$ data above $T_f$  closely follow the resistivity. The fluctuation range $\Ts<T<T_{\rm f}$ and the nematic phase $\TSDW<T<\Ts$ are shaded green and magenta, respectively. From \cite{Kretzschmar:2016} with permission.
  }
  \label{fig:Kretzschmar}
\end{figure*}

Kretzschmar \textit{et al.} \cite{Kretzschmar:2016} focused their attention on  BFCA at finite doping where the magnetic ordering temperature and the structural transition are separated ($T_{\rm SDW}<T_{\rm S}$) as shown in Figure~\ref{fig:Kretzschmar}. The memory function method \cite{Opel:2000} was used for extracting the static Raman relaxation rates $\Gamma_0(T)$ in $A_{1g}$ and $B_{1g}$ symmetry and for facilitating the identification of the cross-over temperature $T_{\rm f}$ below which the contributions from fluctuations become detectable in the $B_{1g}$ channel.

The fluctuations and the particle-hole continuum can only be considered additive if they get excited through independent scattering channels such as, e.g., phonons and charge excitations. If two fluctuations are excited simultaneously (AL mechanism) the response from fluctuations exists independent of other excitations, otherwise the excitations are entangled. In order to separate the fluctuation contribution, the high temperature  particle-hole continuum ($T>T_{\rm f}$) was extrapolated down to low temperatures by varying $\Gamma_0(T)$ in a way that the initial slope of the spectra matches the transport results and was subtracted from the respective total Raman response.

It is considered the key observation of that paper that the fluctuations do not disappear directly below \Ts but continuously lose spectral weight with the peak maximum staying pinned. This indicates a nearly constant correlation length between \Ts and \TSDW. The persistence of the fluctuations down to \TSDW and their immediate disappearance below \TSDW favours their magnetic origin.

The spectral shape of the fluctuations can be described quantitatively in terms AL-type of diagrams if electron-phonon interaction is included \cite{Caprara:2005}. Otherwise the decay on the high-energy side is too slow \cite{Khodas:2015,Karahasanovic:2015}. The corresponding selection rules (see section \ref{sec:selection rules}) yield (${\bf q}_c = \pi$,0) as the wave vector of the critical fluctuations. Furthermore, it was demonstrated that the initial slope of $\chi^{\prime\prime}_{\rm fluct}$ exhibits qualitative agreement with the temperature dependence of the nematic susceptibility in the tetragonal and the nematic phase, as expected from a Ginzburg-Landau type of consideration. A comparison of BFCA and BKFA  and details of the analysis are further discussed in Ref.~\cite{Bohm:2016}.

In FeSe, having a structural transition at $\Ts\approx 90$\,K and short-ranged but no long-ranged magnetic order above 4\,K \cite{WangQS:2016,Baek:2014}, the fluctuations survive down to temperatures right above $\Tc\approx 9$\,K \cite{Massat:2016,Zhang:2017s,Baum:2019}. While the experimental results agree by and large, the interpretations do not. Baum \textit{et al.} \cite{Baum:2019} argue that the persistence of the fluctuations well below \Ts argues for spin fluctuations similar to those in BFCA, whilst Massat \textit{et al.} \cite{Massat:2016} and Zhang \textit{et al.} \cite{Zhang:2017s} assign the fluctuations to charge and/or orbital physics.

Massat and coworkers apply an analysis similar to that in BFCA for studying FeSe at ambient \cite{Massat:2016} and applied pressure \cite{Massat:2018}, as shown in Figure~\ref{fig:Massat}. The spectra exhibit a pronounced temperature dependence in the $B_{1g}$ channel [Figure~\ref{fig:Massat}(a)-(c)]. The static Raman susceptibility $\chi^{\prime}_{B1g}(\Omega= 0, T)$ follows a Curie-Weiss law with $T_0$ significantly lower than $T_{\rm s}$ for reasons discussed above in agreement with the stiffness data. With increasing pressure [Fig.~\ref{fig:Massat}(e)] $T_{\rm s}$ decreases, and above 2\,GPa the SDW is the dominating phase with $T_{\rm SDW}$ reaching 45\,K at approximately 5\,GPa [Figure~\ref{fig:Massat}(d)] \cite{Sun:2016}. With the appearance of the SDW phase the Curie-Weiss-type variation of $\chi^{\prime}_{B1g}(\Omega= 0, T)$ disappears as opposed to the observations in the pnictides [Figure~\ref{fig:Massat}(e) and (f)].

\subsection{Origin of the excitations: spin or charge?}
\label{sec:origin}
There is no consensus yet on the leading instability in the FeBCs, and there are arguments in favor of both orbital and spin ordering. This controversy characterizes also the interpretation of the Raman data, in particular of the fluctuation response. It is true that photons couple to the charge, and if the vertex has the same symmetry as the fluctuations, there is coupling. However, this does not exclude other types of fluctuations to couple to the light; in other words, the selection rules and the coupling argument are not sufficient for deciding between spin and charge. Coupling to the charge does not mean that the photons couple directly  to low-energy charge excitations. Rather, the effective scattering Hamiltonian for particle-hole excitations close to $E_{\rm F}$ is an approximation derived for photon energies much smaller than the gaps in the band structure \cite{Devereaux:2007}. Why this approximation works quite well in the FeBCs and  in the cuprates in the presence of low-lying bands is not entirely clear. It may have its origin in strong correlation effects which broaden all electronic states away from $E_{\rm F}$ and thus reduce resonances.

Selection rules similar to those of charge excitations can also be derived for spin fluctuations owing to the specific band structure of the FeBCs \cite{Kretzschmar:2016,Caprara:2005}. Given the issue with the selection rules, the temperature dependence may be used as another criterion, since critical fluctuations are expected to disappear immediately at the corresponding ordering temperature. This disappearance is observed directly at $\Ts \approx \TSDW$ in BFA \cite{Gallais:2016,Kretzschmar:2016}, at $\TSDW<\Ts$ in BFCA \cite{Kretzschmar:2016}, and not at all in FeSe without long-ranged order \cite{Massat:2016,Baum:2019}. $\chi^{\prime}_{B1g}(\Omega= 0, T)$ as an energy-integrated quantity possibly obscures this important detail to some extent. Therefore, we argue for spin fluctuations to dominate the low-energy Raman spectra in most of the FeBCs but would not go so far as to say that there is evidence beyond reasonable doubt. However, we believe that the issue can be settled, presumably by light scattering, which provides the most direct access to the fluctuations.

The second open question pertains to the nature of the short- or long-ranged spin ordering (which exists beyond any doubt): In the pnictides the spins of itinerant electrons order because of a Fermi surface instability \cite{Mazin:2008,Cruz:2008,Chu:2009}, whereas the spins or at least some of the spins in the chalcogenides are more localized than in the pnictides \cite{Yin:2011,YiM:2015}. The signatures of these different types of order in the Raman spectra are relatively clear as can be directly seen by comparing the Raman response in cuprates to that in Ba122, for instance. However, it was argued that the band at 500\,cm$^{-1}$ originates from incoherent intra-band scattering \cite{W.L.Zhang:2016,Massat:2016} rather than spin excitations \cite{Baum:2019}. Yet, there are two considerations which are hard to reconcile with charge (particle-hole) excitations: (i) Since the energy of the band does not vary significantly upon cooling from 300 to 4\,K, only momentum scattering having a scattering rate of approximately 500\,cm$^{-1}$ and a temperature independent intensity \cite{Zawadowski:1990} can be at its origin. (ii) With the scattering rate being an order of magnitude larger than the superconducting gap the gap excitations would be entirely suppressed \cite{Devereaux:1992,Devereaux:1993} as opposed to the experimental results \cite{Massat:2016,Zhang:2017s,Baum:2019}.

On the other hand, the semi-quantitative agreement of the spectra in all symmetries with simulations \cite{Baum:2019,Ruiz:2019} provides an attractive explanation for the observed spectra. While the dependence (or independence)of the \Blg peak on S substitution \cite{Zhang:2017s} still needs to be clarified it does not seem to be a killer argument \cite{Coldea:2019}. The orbital dependent localization \cite{Yin:2011,Si:2016} provides the necessary explanation for the differences between the pnictides and chalcogenides, including the experimentally observed near localization of the electrons in the $d_{xy}$ orbitals in FeSe as opposed to the by and large orbital-independent itineracy of the electrons in all bands in the pnictides \cite{YiM:2017}.

\section{Superconductivity}
The identification of the pairing mechanism remains one of the major challenges in all unconventional superconductors. The momentum dependence of the gap magnitude (and phase, if possible) is among the important observables for addressing this question since $\Delta_{\bf k}$ and the pairing potential $V_{{\bf k},{\bf k^\prime}}$ are interrelated via the BCS gap equation \cite{Bardeen:1957a}. Therefore, a large variety of methods has been applied to derive properties of the energy gap in the iron-based materials with the goal to understand $V_{{\bf k},{\bf k^\prime}}$ and the coupling \cite{Hirschfeld:2016}. Raman spectroscopy allows access to the magnitude of the gap \cite{Abrikosov:1961,Klein:1984,Devereaux:1994}, its number-phase fluctuations \cite{Leggett:1966,Blumberg:2007,Klein:2010} and bound in-gap states encoding the anisotropy of $V_{{\bf k},{\bf k^\prime}}$ \cite{Monien:1990,Chubukov:2009,Scalapino:2009,Bardasis:1961}.

\subsection{Gap spectroscopy}\label{sec:gap spectroscopy}

\subsubsection{Results from canonical methods}
From the beginning, a substantial band and momentum dependence of the energy gap was observed. Yet, sign changes of the gap on individual bands or between the bands could not be pinned down unambiguously although tunneling experiments with an applied magnetic field \cite{Hanaguri:2010} or neutron scattering experiments \cite{Christianson:2008} supported unconventional order parameters in Fe(Se,Te) and BKFA, respectively. More pieces need to be added to solve the puzzle and clarify the type of pairing in the ground state.

\begin{table*}
  \caption{\label{tab:gap}Compilation of gap energies as observed by angle-resolved photoemission (ARPES),  tunneling (STS), optical (IR) and Raman spectroscopy. For selected cases we show also the results of thermodynamic measurements ($c_{\rm el}$) and neutron scattering (INS) revealing local gap minima and, respectively, the spin resonance energy which are typically below the maximal gap.In the third last column the maximal gap of a method is given in units of $k_{\rm B}\Tc$. A more complete compilation of the results and a discussion may be found in Refs.~\cite{Korshunov:2018} and \cite{Korshunov:2017}. In particular in the case of ARPES but also for some of the thermodynamic and Raman measurements the entries are not exhaustive and display only typical values. In some Raman studies the data of \Blg symmetry were fitted using one, two or three gap scenarios as indicated. For four samples of BKFA the data of all symmetries were fitted simultaneously using a phenomenology including collective modes \cite{Bohm:2014,Bohm:2018}. In some cases the peak frequencies alone are reproduced. For the analysis of the IR spectra the theory by Mattis and Bardeen \cite{Mattis:1958} was applied. The BFCA sample used for STS \cite{Yin:2009} had a nominal doping of $x=0.10$ but the \Tc is more compatible with optimal doping.
  }
  \centering
  \footnotesize
\begin{tabular}{@{}cccccccc}
  \br
  Material&\Tc [K]&Method&$2\Delta_i$ [meV]&2$\Delta_i$ [cm$^{-1}$] & $2\Delta_{\rm max}~[k_{\rm B}$\Tc] &Ref.& comment\\

  \mr
  {Ba(Fe$_{1-x}$Co$_{x}$)$_2$As$_2$} & & & & & & &\\

  $x=0.051$ & 19  &ARPES &8.0/11.6         &                        & 7.1      &\cite{WangM:2016} &el/h band \\

  $x=0.051$ & 18  &Raman &3.7\dots7.4/9.9& 40\dots60/80       & 6.4      &\cite{Bohm:2016} &\Blg/\Alg peaks\\

  $x=0.052$ & 20  &$c_{\rm el}$&3.1/7.1    &                        &4.1     &\cite{Hardy:2010} &min/max gap\\

  $x=0.055$ & 20.5&Raman &5.0              & 41                     &       &\cite{Chauviere:2010} &\Blg; 1-gap\\

  $x=0.055$ & 23  & Raman&8.7/12.4         & 65/105                 & 5.9&\cite{Bohm:2016} &\Blg/\Alg peaks\\

  $x=0.060$ & 24  & Raman&4.1/9.3          & 33/75             &4.5&\cite{Chauviere:2010}&\Blg; 2-gaps\\

  $x=0.061$ & 24  & Raman&8.7/12.4         & 70/100        &5.9&\cite{Muschler:2009}&\Blg/\Alg; aniso.\\

  $x=0.060$ & 25  &$c_{\rm el}$&4.3/10.8   &                        &5.0&\cite{Hardy:2010}&min/max gap\\

  $x=0.065$ & 24.5& Raman&1.9/8.9          & 15/72            &4.2&\cite{Chauviere:2010} &\Blg; 2-gaps\\

  $x=0.067$ & 25  & INS  &$8.0\pm1$        &                        &          &\cite{WangM:2016} &spin resonance\\
  $x=0.067$ & 25  & ARPES& 9.2/13.0        &                        & 6.0&\cite{WangM:2016} &el/h band\\

  $x=0.075$ & 25.5&ARPES &10/13.8          &          &6.2&\cite{Terashima:2009} &el/h band; isotropic\\

  $x=0.075$ & 23.5&Raman & 9.7             & 78                &4.8&\cite{Chauviere:2010} &\Blg; 1-gap\\

  $x=0.085$ & 23  &$c_{\rm el}$&4.0/8.7    &                       &4.4&\cite{Hardy:2010} &min/max gap\\

  $x=0.085$ & 22  &Raman &9.3/11.8         & 75/95          &6.2 &\cite{Muschler:2009}&\Blg/\Alg peaks\\

  $x=0.095$ & 19  &ARPES &9.2/11.2         &                        &6.8&\cite{WangM:2016} &el/h band\\

  $x=0.100$ & 25.3&STS   & $12.5\pm3.0$    &                        &$5.7$ &\cite{Yin:2009} &average; OPT?\\

  $x=0.100$ & 20  &Raman & 9.8             & 79                &5.7&\cite{Chauviere:2010} &\Blg; 1-gap\\
  \mr
  ${\rm Ba_{1-x}K_{x}Fe_2As_2}$  & & & & & & &\\

  $x=0.22$  & 24.6&Raman & 3.1\dots13.7    & 25\dots110       &6.5&\cite{Bohm:2018} &only \Blg peaks\\

  $x=0.25$  & 26  &ARPES & 7.8/15.7        &                    &7.0&\cite{Nakayama:2011} &el/h bands\\
  
  $x=0.25$  & 31  &$c_{\rm el}$&1.9/16.0   &                       &6.0&\cite{Hardy:2016} &min/max gap\\

  $x=0.25$  & 31  &Raman &6.0, 20.0        & 48, 161                &7.5&\cite{WuSF:2017}& \Blg peaks\\
  
  $x=0.25$  & 30.9&Raman &7.4, 18.6        & 60, $150\pm20$         &7.0&\cite{Bohm:2018}& \Blg peaks\\

  $x=0.27$  & 31  & IR   &13.7/27.1        & 110/218              &10.1&\cite{Xu:2017}&Mattis-Bardeen\\

  $x=0.35$  & 38.9&Raman &9.9\dots31.8     & 80\dots256           &9.5&\cite{Bohm:2018}& phenomenology\\

  $x=0.40$  & 38  & INS  &14               &                & &\cite{Christianson:2008}&spin resonance\\
  
  $x=0.40$  & 38  &ARPES &12/24            &                        &7.5&\cite{Ding:2008} &min/max gap\\
  
  $x=0.40$  & 38  &ARPES &8/24             &                 &7.5&\cite{Zhang:2010}&min/max gap; $k_z$\\
  
  
  $x=0.40$  & 38  &ARPES &7.2/20.4         &           &6.2&\cite{Evtushinsky:2014}&min/max gap; $k_z$\\
  
  $x=0.40$  & 37  &STS   &$30$             &                       &$\approx9.4$  &\cite{Wray:2008} &\\
  
  $x=0.40$  & 39.0& IR   &18.6/31.0   & 157/263                  &$9.7$ &\cite{Xu:2017}&Mattis-Bardeen\\

  $x=0.40$  & 38.5&Raman &8.4\dots 32.0    &68\dots258     &9.6 &\cite{Kretzschmar:2013,Bohm:2014}& phenomenology\\

  $x=0.42$  & 38.5&$c_{\rm el}$&2.3/24.2   &                       &7.3&\cite{Hardy:2016} &min/max gap\\
  $x=0.43$  & 36.7&Raman &6.2\dots 31.0     &50\dots 250         &9.8 &\cite{Bohm:2018} &phenomenology\\

  $x=0.48$  & 34.3&Raman &4.0\dots 20.0     &32\dots 161         &6.8&\cite{Bohm:2018}& phenomenology\\

  $x=0.51$  & 34.2&$c_{\rm el}$&2.1/17.7 &                       &6.0&\cite{Hardy:2016} &min/max gap\\

  $x=0.62$  & 26.6&Raman &7.4\dots13.7      &$60\dots110\pm10$   &6.0&\cite{Bohm:2018} &peak energies\\

  $x=0.70$  & 22 &$c_{\rm el}$&2.1/9.1 &                         &4.8&\cite{Hardy:2016} &min/max gap\\

  $x=0.70$  & 21.6&Raman &6.2\dots 11.2     &$50\dots90\pm10$    &6.0&\cite{Bohm:2018} &peak energies\\
  \mr
  {CaKFe$_4$As$_4$}   & {35}  & ARPES  & 16/24 &                     &8.0 &\cite{Mou:2016} &el/h band\\

  {CaKFe$_4$As$_4$}   & {35}  & Raman  & 15.5/26.7  & 125/215 & 8.8 &\cite{Jost:2018} &all symmetries\\
  
  {CaKFe$_4$As$_4$}   & {35}  & Raman &13.6/16.8/20.2&110/135/162&6.7&\cite{ZhangWL:2018}&\Blg; 3-gaps\\
  \mr
  BaFe$_2$(As$_{0.65}$P$_{0.35}$)$_2$&30&ARPES     &4  &          &3&\cite{Shimojima:2011} &isotropic\\
  BaFe$_2$(As$_{0.7}$P$_{0.3}$)$_2$  &30&ARPES &12/15.2&          &5.9&\cite{ZhangY:2012} &nodal gap\\
  BaFe$_2$(As$_{0.5}$P$_{0.5}$)$_2$  &18.2&$c_{\rm el}$&4.8&      &3.1&\cite{Diao:2016} &average gap\\

  BaFe$_2$(P$_{0.5}$As$_{0.5}$)$_2$  &16& Raman    & 6.7      & 54   &4.9   &\cite{WuSF:2016a} & \Blg\\
  \mr
  FeSe      & 9   & STS  & 5.0/7.0          &                        & 9.0     &\cite{Kasahara:2015} &\\
  FeSe      & 8.5 &Raman & 3.6/4.7          & 29\dots38              & 6.4  &\cite{Massat:2016} & \Blg\\
  FeSe      & 8.9 &Raman & 3.0\dots4.6      & 24\dots37      & 6.0 &\cite{Zhang:2017s,Baum:2019}&\Blg\\

  \br
\end{tabular}\\

\end{table*}
\normalsize

Further complication arises since the gaps vary strongly between the families and with elemental substitution \cite{Thomale:2011,Thomale:2009,Thomale:2011a}. In BKFA at optimal doping nearly band dependent gaps with little variation on the individual Fermi surface are observed by angle-resolved photoemission spectroscopy (ARPES) \cite{Ding:2008,Evtushinsky:2009}, while the gaps vary rather strongly around the electron and hole pockets and along the $k_z$ reciprocal axis in overdoped BKFA and BFCA \cite{Hardy:2010,Hardy:2016,Tanatar:2009,Tanatar:2010}. As a general feature, the maximal gaps observed across the families are in the range 6-8 in units of $k_{\rm B}T_{\rm c}$ similar to those in cuprates.

In Table~\ref{tab:gap} we compile characteristic results found by ARPES, specific heat ($c_{\rm el}$), neutron scattering (INS) and  tunneling spectroscopy (STS) and compare them to Raman results. It is instructive to have this overview in addition to the light scattering results which are usually more subject to controversy than the canonical methods, since data analysis is more difficult due to the appearance of additional electronic excitations close to or below the gap energy (c.f. section \ref{sec:higher-order}).

\subsubsection{Raman results}

\begin{figure}
  \centering
  \includegraphics [width=7.0cm]{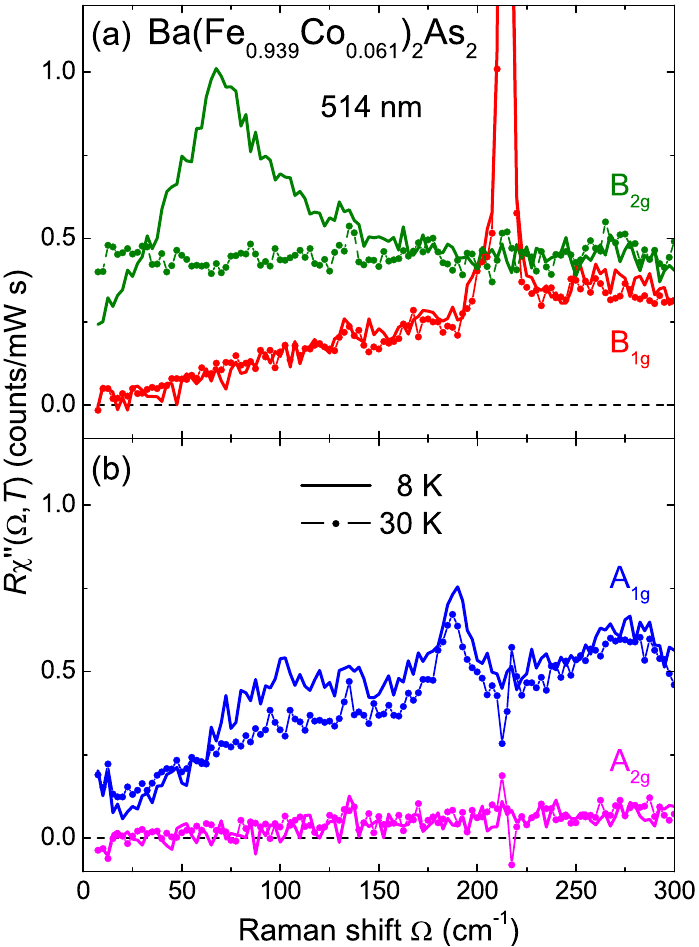}
  \caption[]{(Color online) Symmetry-resolved Raman response of ${\rm Ba(Fe_{1-x}Co_{x})_2As_2}$ ($x=0.061$) for in-plane light polarizations. (a) Here, the 2\,Fe unit cell was used for the symmetry assignment implying that the out-of-phase Fe phonon at 214\,cm$^{-1}$ is observed in the proper $B_{1g}$ symmetry. (b) In $A_{1g}$ symmetry there is a small increase for $\Omega\to0$ from insufficient rejection of the laser light. The $A_{2g}$ signal can safely be neglected at low energies.  From~\cite{Muschler:2009} with permission.
  }
  \label{fig:Muschler}
\end{figure}

Momentum- and band-resolved Raman results for the energy gap were first presented and discussed by Muschler \textit{et al.} \cite{Muschler:2009} and are shown in Fig.~\ref{fig:Muschler}. They reported data of a complete symmetry analysis in the normal and superconducting states of optimally doped ${\rm Ba(Fe_{1-x}Co_{x})_2As_2}$ ($x=0.061$). Here the 2\,Fe unit cell was used, and the related projections in the Brillouin zone were discussed subsequently on the basis of an LDA band structure \cite{Mazin:2010a}. For comparison with Fig.~\ref{fig:srules} the \Blg and \BZg symmetries must be interchanged. The effect of superconductivity is best seen by comparing spectra taken well below and slightly above \Tc similarly as in the case of an SDW (see Fig.~\ref{fig:Chauviere:2011}).

In \BFCA, differences between the spectra above and below \Tc which are associated with the opening of the superconducting gap are observed only in $A_{1g}$ and $B_{1g}$ symmetry. In $A_{2g}$ the intensity is generally low for the absence of excitations having the right symmetry. In $B_{2g}$ symmetry little spectral weight is expected since the vertices are small close to the Fermi surface crossings of the bands. Surprisingly, the intensity becomes comparable to that in $A_{1g}$ and $B_{1g}$ symmetry at approximately 300\,cm$^{-1}$ but there are no indications of an energy gap. There are essentially three explanations for the absence of gap structures: (i) The intensity originates from excitations other than electron-hole pairs such as spins. 
The slow increase of the $B_{2g}$ intensity indicates either that (ii) states far away from the Fermi surface are projected consistent with the vertices or that (iii) the relaxation rate is much higher than expected from transport. Such an anisotropy may occur if the quasi-particle relaxation is strongly momentum dependent for the presence of Co scatterers in the Fe plane. Then the gap excitations are suppressed for specific symmetry projections \cite{Devereaux:1995}. {We consider (iii) most likely since spin excitations are already very weak at optimal doping and gap features indeed appear in $B_{2g}$ symmetry in \BKFA where the substitution generates only out-of-plane defects \cite{Kretzschmar:2013}.}

In superconducting BFCA \cite{Muschler:2009}, the broad maximum close to 100\,cm$^{-1}$ and the well-defined peak at around 70\,cm$^{-1}$ in $A_{1g}$ and $B_{1g}$ symmetry, respectively, correspond to gap excitations in the hole and the electron band. Finite spectral weight observed down to very small energies indicates vanishingly small gaps. The $\sqrt{\Omega}$ dependence in $B_{1g}$ symmetry suggests accidental nodes \cite{Hirschfeld:2009}.

Similar spectra were reported for doping levels in the range $0.045<x<0.10$ \cite{Chauviere:2010,Sugai:2010,Bohm:2016,Muschler:2012} where broad pair-breaking peaks appear in $A_{1g}$ symmetry between 50\,cm$^{-1}$ and 160\,cm$^{-1}$. The $B_{1g}$ spectra generally peak at lower energy than those in $A_{1g}$ symmetry. In both symmetries the spectral changes upon entering the superconducting state become less pronounced below and above optimal doping  \cite{Chauviere:2010,Bohm:2016,Muschler:2012}. In $A_{1g}$ symmetry the peak maxima scale approximately as $6\,k_{\rm B}T_{\rm c}$ in agreement with $2\Delta_{\rm max}$ from other methods. The $B_{1g}$ spectra peak at $4\,k_{\rm B}T_{\rm c}$ at optimal doping and at $3\,k_{\rm B}T_{\rm c}$ for $x<0.055$.

The discussion about the details of the interpretation of the spectral shape in superconducting BFCA is not finally settled although the data agree and the basic features are clear. Chauvi\`{e}re \textit{et al.} \cite{Chauviere:2010} interpret their $B_{1g}$ results for optimally doped and underdoped samples in terms of a two-gap scenario with a maximal gap $2\Delta_{\rm max}\approx 75$\,cm$^{-1}$ and a small but finite gap $2\Delta_{\rm min}\approx 15$\,cm$^{-1}$ (see Table~\ref{tab:gap}). The maximal superconducting gap appears in the same location in momentum space as the SDW gap and is thus suppressed rapidly below optimal doping by the opening of the larger SDW gap, $2\Delta_{\rm SDW}>2\Delta_{\rm max}$, and only the feature related to $2\Delta_{\rm min}$ survives. The spectral shape in $B_{1g}$ symmetry may also be reproduced with a strong $k_z$ dependence of $\Delta_{\bf k}$ but the disappearance of the structure at $2\Delta_{\rm max}$ in underdoped samples can only be explained with an in-plane anisotropy.

Muschler \textit{et al.} \cite{Muschler:2009} argue that the $\sqrt{\Omega}$ variation of the low-energy $B_{1g}$ spectra can be explained with accidental nodes or near-nodes on the outer electron band. This would be in agreement with the results from heat transport \cite{Tanatar:2010a} and theoretical considerations \cite{Mazin:2010a}. Sugai \textit{et al.} find support from band structure calculations for the $B_{1g}$ intensity (1\,Fe cell) to originate from the hole bands \cite{Sugai:2010}. The latter conclusion is at variance with symmetry arguments (cf. Fig.~\ref{fig:srules}) and the LDA results of Mazin \textit{et al.} \cite{Mazin:2010a}. As a matter of fact, the gap in BFCA is very anisotropic at all doping levels and may even vanish for certain momenta. The maximum is in the range $2\Delta_{\rm max}\approx6\,k_{\rm B}T_{\rm c}$. Below optimal doping there is an interaction with the SDW gap.

\begin{figure*}
  \centering
  \includegraphics[width=16cm]{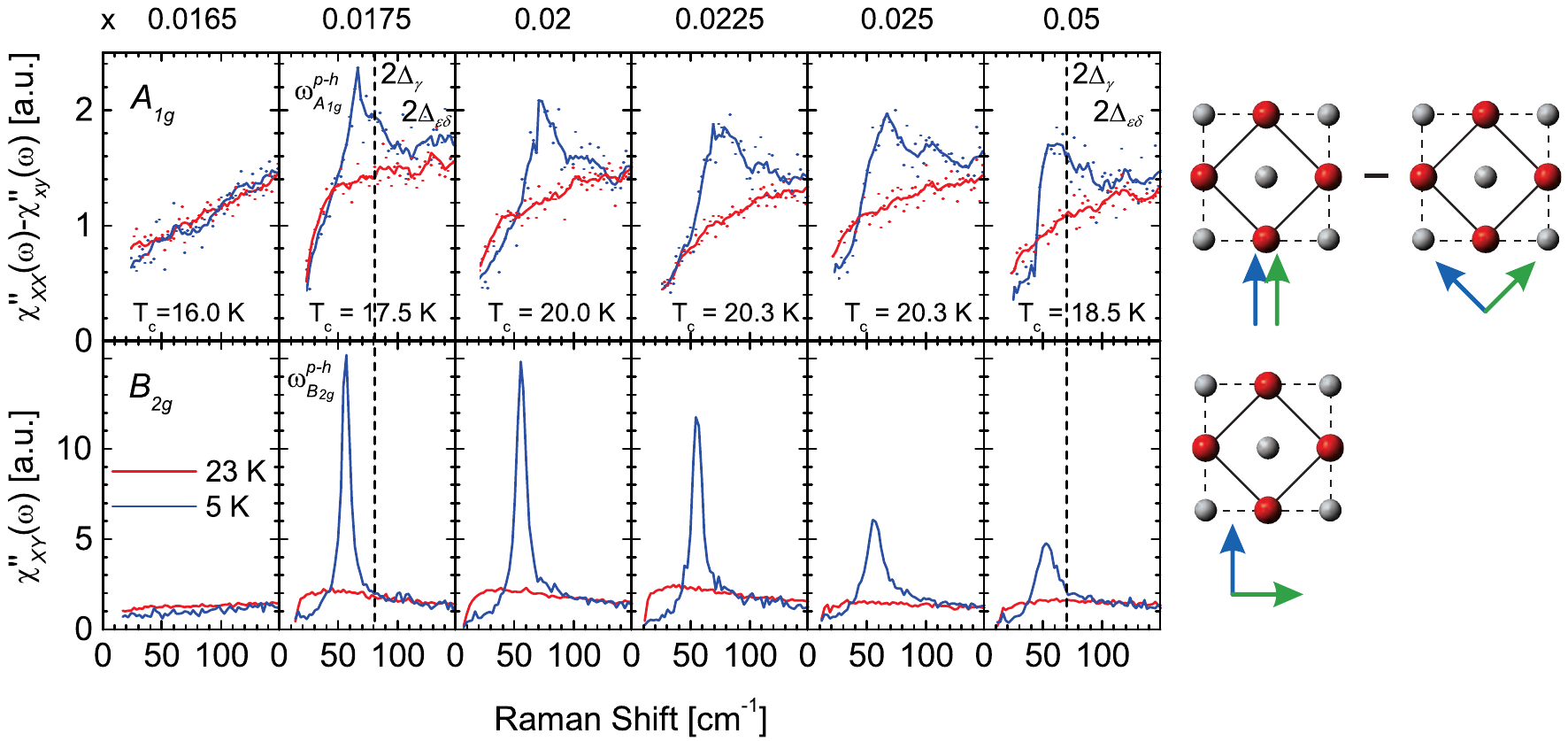}
  \caption[]{(Color online). Raman susceptibilities $\chi^{\prime\prime}_{XX}(\omega)$-$\chi^{\prime\prime}_{xy}(\omega)$ and $\chi^{\prime\prime}_{XY}(\omega)$ in the superconducting state for excitation in the blue (476\,nm). From \cite{Thorsmolle:2016}. The polarizations are indicated on the right by NL and RH. (a) $\chi^{\prime\prime}_{XX}(\omega)$-$\chi^{\prime\prime}_{xy}(\omega)$ (top row) and $\chi^{\prime\prime}_{XY}(\omega)$ (bottom row) in the superconducting (5\,K) and normal (23\,K) states at doping levels as indicated. The vertical dashed line, shown for $x=0.0175$ and 0.05, indicates the lowest superconducting gaps $2\Delta_\gamma\simeq10$\,meV and 9\,meV, respectively, determined by ARPES. $B_{2g}$ symmetry in the figure corresponds to $B_{1g}$ used in this review.
  }
  \label{fig:Thorsmolle}
\end{figure*}

The observation of gap features in other pnictides and in chalcogenides was delayed by sample and surface issues. In chalcogenides the observation of pair breaking succeeded first in Rb$_{0.8}$Fe$_{1.6}$Se$_2$ after cleaving the sample \textit{in situ} at low temperature \cite{Kretzschmar:2013}. Differences between normal and superconducting spectra were only found in $B_{1g}$ symmetry. Since Rb$_{0.8}$Fe$_{1.6}$Se$_2$ has presumably no hole pockets, the $B_{1g}$ selection rule supports the symmetry assignment of Refs. \cite{Muschler:2009} and \cite{Mazin:2010a} as reproduced in section \ref{sec:selection rules}. The observed gap appears to be almost constant on the electron pockets, being compatible with either a simple $s$- and a $d$-wave state without nodes on the Fermi surface or a sign change between the pockets. From what we shall see below $d$-wave symmetry is more likely but cannot be distinguished from an $s$-wave gap on the basis of the light scattering experiment. As in the case of BFCA, it is difficult to explain the intensities in other symmetries which do not show features induced by superconductivity. As a hypothesis which needs to be worked out in more detail we assign the continua in $A_{1g}$ and $B_{2g}$ symmetry to excitations in lower lying bands and/or the tails of the numerous phonon lines.

In ${\rm Ba_{0.6}K_{0.4}Fe_{2}As_2}$ continua and gap structures are observed in all symmetries and are by and large in agreement with the results from other methods \cite{Kretzschmar:2013} as shown in Table~\ref{tab:gap}. Since the hole-like Fermi surfaces are more extended in the Brillouin zone, the relevant electronic states may be sampled by all vertices. Additional support comes from simulations using the effective mass approximation and leading to a semi-quantitative explanation of the spectra \cite{Bohm:2014}. Yet, resonance effects may also contribute \cite{Bohm:2018,Thorsmolle:2016,Baum:2018} although they were found to be mild in BFCA upon exciting with blue and green light \cite{Mazin:2010a}.

The spectra of ${\rm Ba_{0.6}K_{0.4}Fe_{2}As_2}$ clearly show a small but true gap of approximately 30\,cm$^{-1}$ indicating, among other things, that the contributions from luminescence are negligible here. The largest gaps reside on the middle hole band and on the electron bands and are rather sharp as opposed to BFCA. The gap maximum $2\Delta_{\rm max}$ reaches almost 10 in units of $k_{\rm B}T_{\rm c}$ \cite{Kretzschmar:2013,Bohm:2014}. This large ratio exceeds the ARPES results and, in particular, the gap ratio derived from the electronic specific heat as a typical bulk method \cite{Hardy:2016}. However, thermally activated behaviour is generally sensitive to the small gaps on the individual bands. These minimal gap energies are typically 180\,cm$^{-1}$ or 6.8 in units of $k_{\rm B}T_{\rm c}$ in the phenomenological analysis \cite{Bohm:2014,Bohm:2018} thus reconciling Raman scattering and thermodynamic measurements.

In addition to the pair-breaking features, the $B_{1g}$ spectra show unexpected structures below $2\Delta_{\rm max}$ which, as opposed to the pair-breaking features, are nearly resolution limited and will be discussed in detail in the following section. Here we first wrap up observations of the energy gaps by Raman scattering in other compounds.

A pair-breaking peak was also observed in the $A_{1g}$ spectrum of ${\rm BaFe_2(As_{0.5}P_{0.5})_2}$ \cite{WuSF:2017} with the spectral weight decreasing linearly towards low frequencies indicating the presence of nodes in agreement with recent ARPES results for optimally doped ${\rm BaFe_2(As_{0.65}P_{0.35})_2}$ \cite{ZhangY:2012} but, at first glance, not with thermodynamic results on a material with comparable doping \cite{Diao:2016} (see Table~\ref{tab:gap}). However, the small value of the thermodynamically derived gap indicates a substantial anisotropy also for ${\rm BaFe_2(As_{0.5}P_{0.5})_2}$ which may be conceiled by the notion of a single (average) gap.

Very recently, pair-breaking features were observed in CaKFe$_4$As$_4$ \cite{Jost:2018,ZhangWL:2018}. Depending on the incoming photon energy, gap features are observed either in all symmetries \cite{Jost:2018} or only  in $B_{1g}$ symmetry (1\,Fe unit cell corresponding to $B_{2g}$ in the 2\,Fe cell used by Zhang \textit{et al.} \cite{ZhangWL:2018}). The gap energies derived from the $A_{1g}$ and $B_{2g}$ (1\,Fe) Raman data are compatible with those determined by ARPES \cite{Mou:2016} if the entire energy range given by Mou \textit{et al.} is considered (see also Table~\ref{tab:gap}). The $B_{1g}$ maximum (1\,Fe) appears at a substantially smaller energy, has a sharp onset and is clearly peaked \cite{Jost:2018}. Zhang \textit{et al.} \cite{ZhangWL:2018} identify all features observed in $B_{1g}$ symmetry with pair breaking while Jost \textit{et al.} \cite{Jost:2018}, based on the results in three symmetries, propose that the $B_{1g}$ peak at 134\,cm$^{-1}$ originates from a collective mode similar to that in slightly underdoped BKFA \cite{Bohm:2018}. This - controversial - point of view would indicate a sub-leading interaction having $d_{x^2-y^2}$ symmetry which is to be discussed in more detail now.

\subsection{Collective modes}
\label{sec:collective}

\begin{figure*}
  \centering
  \includegraphics [width=12cm]{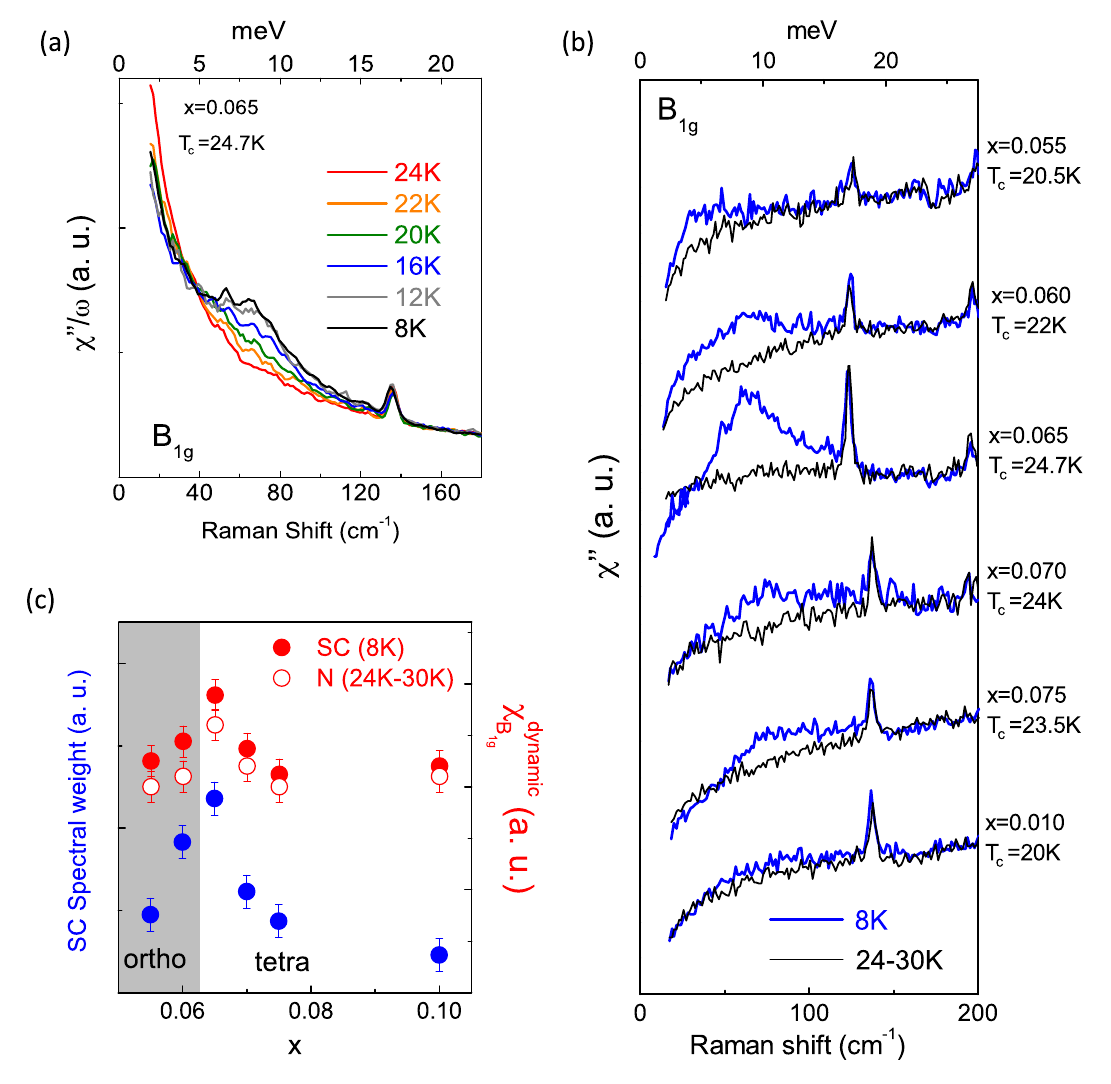}
  \caption[]{(Color online) Doping dependence of the Raman spectra of ${\rm Ba(Fe_{1-x}Co_{x})_2As_2}$. (a) Evolution of the $B_{1g}$ Raman conductivity $\chi''/\omega$ across \Tc for $x=0.065$. (b) $B_{1g}$ Raman response well below (blue) and right above \Tc (black) as a function of Co doping as indicated. (c) Integrated SC spectral weight of the Raman response $\chi''$ (blue) as a function of Co doping. The corresponding nematic susceptibilities $\chi^{\rm dynamic}_{B{1g}}$ both slightly above \Tc  (N) and well below $T_c$ (SC) are also shown (open and, respectively, full red symbols). From \cite{Gallais:2016a}.
  }
  \label{fig:Gallais}
\end{figure*}

The lowest order gap excitations in the electronic Raman spectra essentially reflect the magnitude of the energy gap. Yet, some caution is appropriate when numbers are to be derived (see table~\ref{tab:gap}). Similarly to the ARPES or tunneling spectra neither the onset of the enhanced intensity nor the peak maxima are directly related to the gap $2\Delta$. Only in the clean limit, for $q=0$, and an isotropic superconductor a square-root singularity is expected at $2\Delta$ \cite{Klein:1984}. In all other cases the maximum is at higher energies \cite{Klein:1984,Devereaux:1994,Devereaux:1992,Devereaux:1995}, and numbers can only be extracted \textit{via} theoretical models. Specifically in multiband systems, such as the FeBCs or in the presence of higher order corrections (final state interactions), only a model analysis leads to useful conclusions as outlined in paragraph~\ref{sec:higher-order}. Yet, since a host of additional information can in principle be derived from the spectra in general and from higher order contributions specifically it is worthwhile.

There is general consensus that collective modes exist in at least some of the pnictides having sufficiently clean gaps \cite{Kretzschmar:2013,Bohm:2014,Bohm:2018,Thorsmolle:2016,Jost:2018}. Indications of collective modes were also reported for ${\rm K_{0.75}Fe_{1.75}Se_2}$ and discussed along with a theoretical model \cite{Khodas:2014} but we are not aware of a comprehensive symmetry analysis or an in-depth study. Including this report, collective modes were observed mainly in \Blg (1\,Fe) or \BZg (2\,Fe) symmetry (which are equivalent). The \Alg collective mode, as predicted by Chubukov \textit{et al.} \cite{Chubukov:2009}, was observed as part of a broad spectrum but not as an isolated line.

There is a lively discussion on how the collective modes are to be explained in terms of one of the essentially three possibilities (see section \ref{sec:higher-order}): (i) Leggett modes, (ii) fluctuation modes which become undamped in the presence of a gap and (iii) BS modes. Here, the distinction between particle-particle and particle-hole modes \cite{Thorsmolle:2016} is somewhat artificial, and both of them were actually coined excitons in the original paper of Bardasis and Schrieffer \cite{Bardasis:1961}. We adopt this nomenclature in the following. The essential difference is that particle-particle and particle-hole bound states are expected for attractive and, respectively, repulsive contributions to an attractive pairing potential and \textit{vice versa}. From an experimental point of view a distinction is difficult or impossible.

(i) Leggett modes \cite{Leggett:1966} were first discussed for MgB$_2$ \cite{Blumberg:2007,Klein:2010} where the intra-band interaction dominates and the Leggett modes are below the gap edge. In FeBCs there is a wide agreement that the intra-band interaction is weaker than the inter-band interaction \cite{Mazin:2008,Boeri:2008}, and the Leggett modes are expected to be pushed towards the continuum and overdamped. They may contribute to the Raman intensity at the gap edge and are thus indistinguishable from the pair-breaking effect \cite{Burnell:2010,Cea:2016}. Consequently, they are unlikely to augment the information derived from gap spectroscopies, although interesting conclusions about the pairing symmetry could be derived in special cases of chalcogenides without a central Fermi surface \cite{Huang:2018}. Here the Leggett modes are predicted to appear in \Blg symmetry, whereas, in the presence of a Fermi surface encircling the $\Gamma$ point, as in all pnictides and in bulk FeSe, the Leggett modes are expected to be observed in \Alg symmetry \cite{Cea:2016}. This argument needs to be qualified if the orbital content of the bands is taken into account \cite{Burnell:2010}.

(ii) In NaFe$_{1-x}$Co$_x$As, when excited with blue light (476\,nm), a very strong and narrow $B_{2g}$ mode ($B_{1g}$ in the 1\,Fe unit cell) appears below \Tc at approximately 56\,cm$^{-1}$ close to the gap edge derived from ARPES \cite{Thorsmolle:2016}. As shown in Figure~\ref{fig:Thorsmolle}, 56\,cm$^{-1}$ is close to the maximum of the fluctuation peak observed above  \Tc. The continuous temperature dependence across \Tc, the narrowing below \Tc and the independence of the mode energy of \Tc support the interpretation in terms of a quadrupolar fluctuation of charges between the electron and hole bands which becomes undamped inside the superconducting gap.

In $A_{1g}$ symmetry a broad peak is observed which cuts off softly below the maximum at approximately 70\,cm$^{-1}$ thus indicating a finite density of states inside the gap. The maximum - as an integal part of the peak - is interpreted in terms of the particle-hole collective Bardasis-Schrieffer mode predicted by Chubukov, Eremin, and Korshunov \cite{Chubukov:2009} for the case of an $s_\pm$ ground state and an $s_{++}$ subleading instability induced by  orbital fluctuations \cite{Kontani:2010}. Since the relatively broad peak includes several excitations, the gap energy can be extracted only with difficulties from the smoothed $A_{1g}$ spectra. For $x=0.0175$ and 0.05 the humps on the high-energy side are close to the ARPES gaps.

The authors also used red photons (646\,nm) for excitation but show only $B_{2g}$ spectra, making a comparison with the spectra obtained for blue light less stringent. For 646\,nm, the $B_{2g}$ excitation at 56\,cm$^{-1}$ becomes much weaker in the underdoped range, $x\le0.0175$, and two new modes appear. These modes are compared to the BS mode in the $A_{1g}$ spectrum measured with excitation at 476\,nm and are tentatively assigned to p-h and p-p collective modes without experimental substantiation or a phenomenological theory. Therefore, more work is needed to disentangle the complex but very interesting Raman spectra of NaFe$_{1-x}$Co$_x$As.

Whereas the normal-state data of NaFe$_{1-x}$Co$_x$As are rather similar to those of BFCA several differences are observed below \Tc. For instance, the gap anisotropy on the individual bands is larger in BFCA than in NaFe$_{1-x}$Co$_x$ as can be inferred from the Raman spectra \cite{Muschler:2009,Chauviere:2010,Thorsmolle:2016} or, similarly, from other experiments \cite{Tanatar:2009,Liu:2011,GeQQ:2013}. Given the rather anisotropic gap in BFCA, it is not entirely surprising that no sharp in-gap modes comparable to those in NaFe$_{1-x}$Co$_x$As are observed. On the other hand, the peak maximum in $B_{1g}$ symmetry is quite sharp in optimally doped BFCA and may be alternatively interpreted in terms of a nematic resonance near a quantum critical point \cite{Gallais:2016a,Gallais:2016}. In BFCA, both the enhancement of the spectral weight of the $B_{1g}$ pair-breaking peak upon approaching optimal doping, $x\approx0.065$, and its scaling with the nematic response above $T_c$ (Figure~\ref{fig:Gallais}) argue in favour of the nematic resonance. Yet, a similar doping dependence is also observed in $A_{1g}$ symmetry and qualifies this conclusion \cite{Bohm:2016}.

Na111 and BFCA seem to be the two material classes with the strongest interaction between superconductivity and nematic fluctuations. In contrast, the fluctuations can hardly be observed in BKFA \cite{Bohm:2016,WuSF:2017} or CKFA \cite{Zhang:2018}, and a detailed comparison of these material classes seems highly desirable.


(iii) Finally, we discuss the possibility of sub-leading pairing interactions having $d_{x^2-y^2}$ symmetry and the related BS modes inside the gap in the \Blg Raman spectra. The BS modes display various properties which distinguish them from other collective modes [see Eq.~(\ref{eq:responsecm})].\\
(a) In a clean gap the BS modes are resolution limited. The energy, $\Omega_{\rm BS}(T)$, is directly linked to the gap parameter \cite{Monien:1990}, as opposed to the maximum of the pair-breaking peak $\Omega_{\rm pb}(T)$ which depends on both the gap $\Delta_{\rm max}(T)$ and the quasi-particle relaxation rate $\Gamma_{\rm qp}(T)$ as $\Omega_{\rm pb}(T)\approx2\sqrt{|\Delta_{\rm max}(T)|^2+\Gamma^2_{\rm qp}(T)}$ \cite{Devereaux:1992,Devereaux:1993,Devereaux:1995,Manske:2004}. Thus the temperature dependence of the BS modes rather than that of $\Omega_{\rm pb}(T)$ is expected to be determined by that of the single particle gap, $\Omega_{\rm BS}\propto \Delta_{\rm max}(T)$. \\
(b) The BS mode drains spectral weight from the pair-breaking peak, but there is no sum rule. Rather, the intensity in the pair-breaking maximum is reduced rapidly, whereas the spectral weight in the BS mode increases first with increasing interaction strength $\lambda_\alpha$, with $\alpha$ indexing the eigenvalues (see section \ref{sec:doping}), and then decreases towards zero \cite{Bohm:2018}. In isotropic systems the intensity in the pair-breaking maximum is reduced in the entire energy range. In systems with anisotropic interactions $V_{{\bf k},{\bf k}^\prime}$ only parts of the pair-breaking peak are depleted depending on the channel-specific components of $V_{{\bf k},{\bf k}^\prime}$. This behavior can be modeled phenomenologically \cite{Bohm:2014} or on the basis of the eigenvectors $g_\alpha({\bf k})$ which determine the momentum dependence of the gap $\Delta_\alpha({\bf k})$ and, to some extent, reflect the variation of $V_{{\bf k},{\bf k}^\prime}$ by virtue of the BCS gap equation \cite{Maiti:2016,Bohm:2018}.\\
(c) The binding energies of the BS modes, $E_{{\rm BS},\alpha} = 2\Delta_{\rm max}-\Omega_{{\rm BS},\alpha}$, are related to the coupling strengths of the sub-leading channels $\lambda_\alpha$ ($\alpha>1$) with respect to that of the ground state $\lambda_1$.  For $\lambda_1\approx1$ the relationship is given by $\sqrt{E_{{\rm BS},\alpha}/2\Delta_{\rm max}}\approx\lambda_\alpha/\lambda_1$ and is thus much simpler than that of the intensities \cite{Bohm:2018}.

\begin{figure}
  \centering
  \includegraphics [width=6.5cm]{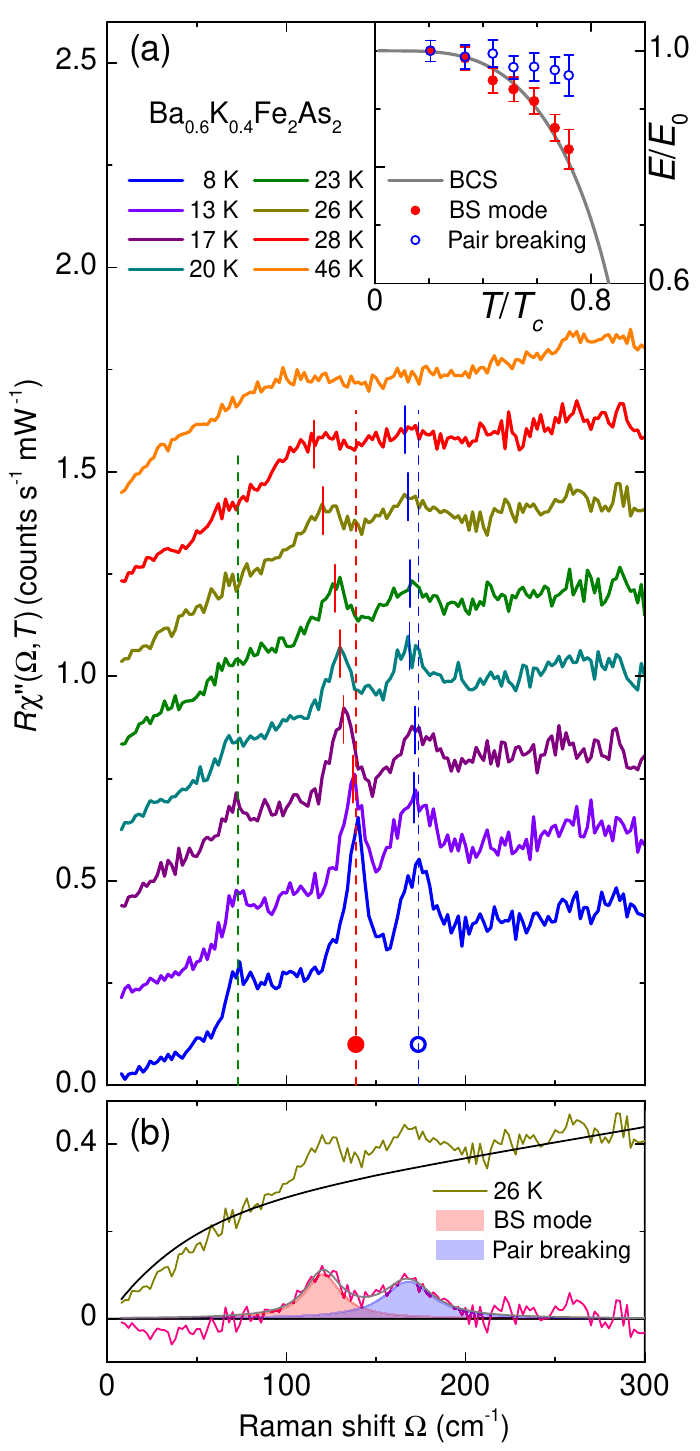}
  \caption[]{(Color online)  Temperature dependence of the Raman spectra of Ba$_{0.6}$K$_{0.4}$Fe$_2$As$_2$ in B$_{1g}$ symmetry. (a) The spectra measured above 8\,K are consecutively shifted up by 0.2 units. The pair-breaking features (open symbols) and the collective mode (full circles) depend differently on temperature, as shown in the inset (where zero energy is suppressed). The pair-breaking maximum exhibits a temperature dependence which is different from the BCS prediction due to interaction effects. (b) The peak energies are determined by fitting the spectra with two Lorentzians and a smooth phenomenological background (black curve). From \cite{Bohm:2014} with .
  }
  \label{fig:Bohm}
\end{figure}


All $B_{1g}$ peaks observed in Ba$_{0.6}$K$_{0.4}$Fe$_{2}$As$_2$ were suggested to be BS modes \cite{Kretzschmar:2013} following the phenomenology for anisotropic gaps proposed by Scalapino and Devereaux \cite{Scalapino:2009}. The detailed study of the temperature dependence performed later \cite{Bohm:2014} is shown in Fig.~\ref{fig:Bohm} and requires this assignment to be revisited. The inset demonstrates the scaling of $\Omega_{\rm BS}(T)$ at 140\,cm$^{-1}$ and $\Delta_{\rm max}(T)$ as directly observed by ARPES \cite{Evtushinsky:2009}, while the maximum at 170\,cm$^{-1}$ stays pinned. Apparently, the two strongest modes depend distinctly differently on temperature suggesting the mode at 170\,cm$^{-1}$ to be related to pair breaking and that at 140\,cm$^{-1}$ to a sub-leading channel. The comparison of all symmetries shows that intensity is in fact drained from the \Blg pair-breaking maximum although part of the peak survives indicating highly anisotropic interactions.

Maiti \textit{et al.} \cite{Maiti:2016} pointed out that there may be more than one BS mode in the presence of a hierarchy of sub-leading coupling channels in addition to the $s_\pm$-wave ground state. Although there is a candidate peak at 70\,cm$^{-1}$ (see Fig.~\ref{fig:Bohm}), this proposal can only be addressed by studying differently doped samples.

\subsection{Doping}\label{sec:doping}
Doping $x$ or pressure $P$ can be used as non-thermal control parameters being relevant in the context of quantum phase transitions. Here, doping $x$ proves useful for the assignment of the in-gap modes and for scrutinizing the anisotropy of the pairing potential $V_{{\bf k},{\bf k}^\prime}$ through the evolution with $x$ of the related in-gap modes in BKFA. This will be the main focus of this subsection, but prior to this discussion the results on BFCA and BFAP will be summarized. For Na111 the reader is referred to section~\ref{sec:collective}.

The parent compound Ba122 can be driven superconducting in various ways. Both chemical substitution, using isovalent phosphorus substitution for arsenic, and applied pressure lead to \Tc values in the 30\,K range \cite{Colombier:2009,Analytis:2014}. Currently there are no Raman studies of pressure-induced superconductivity in BFA. However, BFAP can be considered to fill this gap at least as a proxy \cite{WuSF:2016a}. For $x=0.5$ BFAP has a \Tc of 16\,K and displays a broad pair-breaking peak in \Alg symmetry having a maximum at $\Omega_{\rm pb}(T) \approx 2\Delta_{\rm max}=6.7$\,meV or 4.9\,$k_{\rm B}T_{\rm c}$. The nearly linear energy dependence of the spectra below the peak maximum indicates a much broader gap distribution than in BKFA and suggests line nodes of the gap. If the peak is identified with the gap maximum of 4.9$k_{\rm B}T_{\rm c}$, it falls below the ratio in the range of 6-8\,$k_{\rm B}T_{\rm c}$ for other compounds \cite{Korshunov:2018}. There are no gap structures in the other symmetries and no collective modes in any symmetry. Thus from all aspects BFAP is closer to BFCA than BKFA.

Due to the doping dependent changes of the band strucure, ($\pi$,$\pi$) scattering is expected to gain strength in BFCA in comparison to BKFA, and one would expect enhanced sub-dominant coupling channels. Rather, the anisotropy of the gaps grows, and the resulting density of states below the gap maximum leads to overdamping of potential in-gap modes similar to what has theoretically been shown to happen for $d$-wave gaps \cite{Devereaux:1995a}. Therefore no collective modes can be resolved in BFCA and the doping dependence is limited to intensity variations of the pair-breaking features described above. Consequently, only BKFA and the related CKFA \cite{Jost:2018} facilitate the study of changes of $V_{{\bf k},{\bf k}^\prime}$ as a function of doping or, more appropriately, of the Fermi surface topology.

In what follows we assume that the modes observed below the maximal gap in BKFA are excitonic in origin \cite{Bardasis:1961}. This interpretation is not entirely accepted, although many criteria were tested experimentally (see above) and found to be in agreement with the theoretical prediction whereas counterarguments were not presented yet. The doping dependence adds another piece of evidence to this assignment.

Although BKFA is superconducting for $0.1 < x \le 1$ the range without magnetic order or changes of the Fermi surface topology is rather small, $0.25<x<0.6$. For $x<0.25$ BKFA develops an SDW which gaps out part of the Fermi surface. For $x>0.6$ the Fermi energy dives below the bottom of the inner electron band and for $x>0.7$  hole-like bands appear around the $X$ points \cite{Xu:2013}. It is still under debate which doping should be associated with the Lifshitz transition but, as a matter of fact, one electron band is lost at $x\approx0.6$.

The doping dependence of the Raman spectra in superconducting BKFA was studied by two groups. Wu and coworkers \cite{WuSF:2017} looked at three doping levels, $x=0.25$, 0.4, and 0.6 and reproduced earlier results \cite{Kretzschmar:2013} for $x=0.4$. Three peaks were observed in $B_{1g}$ symmetry at 50, 120, and 168\,cm$^{-1}$ and at 70, 140, and 172\,cm$^{-1}$ for the first and the second cleave, respectively, of one crystal. Sample-dependent differences at optimal doping were also observed by Kretzschmar and collaborators \cite{Kretzschmar:2013} but the variations were much smaller, in particular the peak energies were nearly and the overall intensities entirely identical. We find it difficult to explain that the results obtained from two successive cleaves of the same crystal differ substantially, while the local $T_{\rm c}$ values or doping concentrations $x$ apparently do not. For $x=0.25$ Wu \textit{et al.} observe spectra which are qualitatively different from those at optimal doping and similar to what was found later by B\"ohm \textit{et al.} \cite{Bohm:2018} in comparable samples. The spectra of Wu's overdoped BKFA, having nominally $x=0.6$ and $T_{\rm c}=25$\,K, are closer to the results found for $x=0.43\dots0.48$ in Ref.~\cite{Bohm:2018} although the \Tc values differ by at least 10\,K. An explanation of these discrepancies without directly comparing the magnetization measurements of the samples studied cannot be a subject of this review.

\begin{figure}
  \centering
  \includegraphics [width=8.5cm]{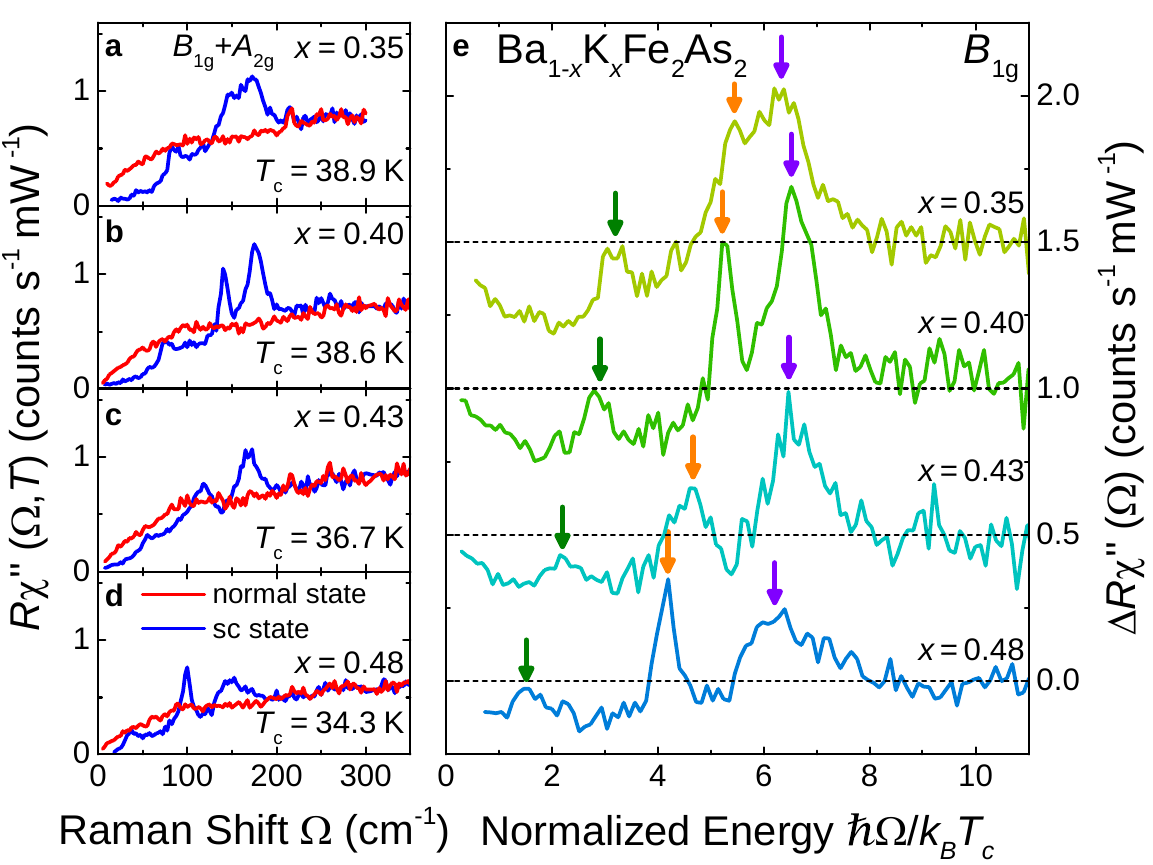}
  \caption[]{\Blg Raman spectra of BKFA for doping levels as indicated. (a)-(d) Raw data (after dividing by the Bose-Einsten factor) slightly above (red) and well below \Tc (blue). (e) Difference spectra $\Delta R \chi^{\prime\prime}(\Omega) = R \chi^{\prime\prime}(\Omega,T\approx8\,{\rm K})-R \chi^{\prime\prime}(\Omega,T\gtrsim\Tc)$. Here all temperature independent features drop out. Apparently, there are no temperature-dependent phonons. As an artifact, the intensity becomes negative inside the gap. The respective zero-intensity lines are indicated (dashes). From \cite{Bohm:2018} with permission.
  }
  \label{fig:B1-doping}
\end{figure}

Regarding the experiments in the doping range $0.35\le x\le0.48$ studied in Ref. \cite{Bohm:2018} only samples without indications of secondary transitions and with $\Delta T_{\rm c}<1.3$\,K were selected. In this relatively small doping range which is sufficiently far away from the SDW and from changes in the Fermi surface topology the Raman spectra of all symmetries depend continuously on $x$. The highest peak energies in all symmetries follow \Tc to within $\pm12$\%. In general, the \Alg and \BZg energies are close to $8\pm1.0\,k_{\rm B}T_{\rm c}$ and thus higher than those in \Blg symmetry which scale roughly as $6.2\,k_{\rm B}T_{\rm c}$ as shown in Fig.~\ref{fig:B1-doping} which displays raw and difference spectra in \Blg symmetry. There are two other maxima in \Blg symmetry at lower energy which are clearly resolved in all data sets and scale as $1-x$ rather than \Tc [Fig.~\ref{fig:B1-doping}(e)]. The comparison of all symmetries and doping levels demonstrates that there are only very weak or no maxima in \Alg and \BZg symmetry in the range of the low-energy \Blg peaks. In addition, the low-energy peaks are nearly resolution limited and depend on temperature as $\Delta(T)$.

For $0.35\le x\le0.48$ the spectra of all symmetries can be described consistently \cite{Bohm:2018} in terms of the phenomenology outlined in Ref.~\cite{Scalapino:2009} and further elaborated and described by B\"ohm \textit{et al.} \cite{Bohm:2014} for $x=0.4$. This approach starts from a realistic electronic structure \cite{Graser:2009}, and the doping is accounted for by a shift of the Fermi energy. The Raman vertices are derived from the band structure [see Eqs.~(\ref{Eq:A1g}), (\ref{Eq:B1g}), and (\ref{Eq:B2g})]. The \Alg and \BZg spectra are used for determining the band and momentum dependent gap values compatible with ARPES studies. After some iterations the bare \Blg spectra (without final state interaction) become consistent with the experimental \Blg spectra. Here, consistent does not mean that the entire \Blg spectra can be reproduced. Rather, there are three features which by no combination of gaps can be explained, the two sharp lines below the gap edge and the missing intensity above the \Blg pair-breaking maximum which is expected from the \Alg and \BZg spectra and should therefore also show up in \Blg symmetry. With the final state interaction ``switched on'' this part of the calculated \Blg spectrum is suppressed and reappears in the narrow modes. An explicit calculation was performed only for the stronger mode which was then found to acquire too much spectral weight for the coupling strength $\lambda_d$ derived from the energy position [see Eq.~(\ref{eq:responsecm})].

This discrepancy was solved later when theoretical considerations suggested the existence of two sub-leading channels rather than one \cite{Maiti:2016,Bohm:2018}. The two sub-leading channels ($\alpha=2,~3$ ordered by strength) were derived from two independent microscopic approaches and were found to have first and second order $d_{x^2-y^2}$ (\Blg) symmetry (see Fig.~\ref{fig:vertices} center). Having the higher binding energy $E_{{\rm BS},2}>E_{{\rm BS},3}$ the intensity of the BS mode at lower absolute energy (higher binding energy) is much smaller but still high enough for being the strongest spectral feature in the respective energy range. Unfortunately, the gap energies of the outer hole bands are in the same range \cite{Hardy:2016}, motivating Wu \textit{et al.} to assign the mode to that gap \cite{WuSF:2016a}. Yet, both the phenomenology and the experimental results in \Alg and \BZg symmetry show that the \Blg mode is at least an order of magnitude too strong for justifying an explanation in terms of direct gap excitations \cite{Bohm:2018} thus furnishing further evidence for the excitonic character of the two narrow in-gap modes.

Very recently, CKFA was studied. CKFA is a stoichiometric version of BKFA since the Ca and K layers alternate in a regular fashion. From the viewpoint of valence count it should be slightly overdoped, and the \Tc values are indeed close to the maximum found for BKFA. The ARPES \cite{Mou:2016} and Raman experiments \cite{Jost:2018,ZhangWL:2018} find gaps similar to those of BKFA. The features induced by superconductivity are relatively strong in \Blg symmetry. In \Alg and \BZg symmetry they are weak and can only be observed for $\hbar\omega_I=2.16$\,eV \cite{Jost:2018} but not for $\hbar\omega_I=1.92$\,eV \cite{ZhangWL:2018}. The weak structures in \Alg and \BZg symmetry are compatible with the gaps derived from ARPES. The \Blg spectra have substructures similar to those found for $x=0.35$ in BKFA. There is agreement that the mode at 134\,cm$^{-1}$ may be a collective excitation and that the hump at 160\,cm$^{-1}$ is a remainder from pair breaking. The very weak structure at 50\,cm$^{-1}$ tentatively assigned to a second BS mode in Ref.~\cite{Jost:2018} remains controversial.

Even though there is no agreement among the experimental groups about the details of the interpretation and, in particular, the doping dependence in BKFA, the question arises as to whether or not the idea of competing pairing channels may be a relevant contribution from Raman scattering to directly support the microscopic considerations.  For addressing this question the hierarchy of pairing interactions was studied.

\subsection{Possible conclusions for Cooper pairing}
\label{sec:pairing}
In conventional superconductors, the ground state has a much lower energy than potential competing pairing tendencies. Unconventional superconductors have typically various instabilities in close proximity, all of which may be intertwined with Cooper pairing. The ways to study the related phase diagrams include Hubbard-like models \cite{Kontani:2010}, the spin-fluctuation scenario which is studied in the random phase approximation (RPA) \cite{Mazin:2008,Maier:2009}, and the functional renormalization group (fRG) scheme \cite{Thomale:2011,Chubukov:2008b,Thomale:2009,Platt:2014} which, as opposed to RPA, treats all possible interactions on equal footing.

\begin{figure}
  \centering
  \includegraphics [width=6.0cm]{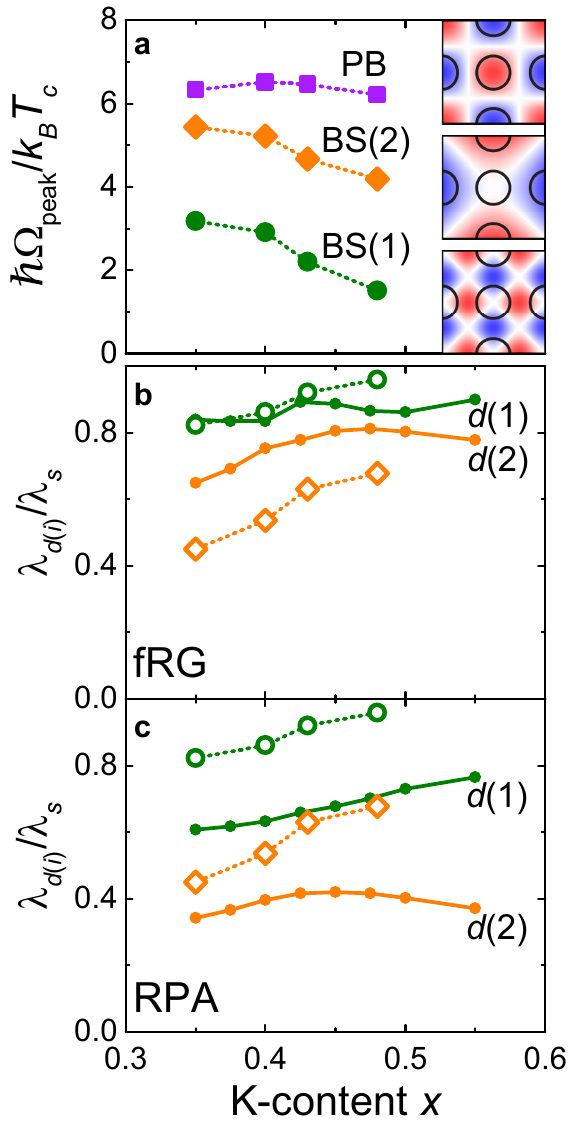}
  \caption[]{Doping dependence of the pairing strength in BKFA.  (a) The positions of the pair-breaking maxima scale approximately as \Tc whereas the energies of the BS modes decrease monotonously with increasing doping in the range $0.35<x0.48$ indicating increasing coupling in the sub-leading channels. The insets show the eigenvectors $g_\alpha({\bf k})$ of the tree channels. (b) and (c) The coupling strength of the two sub-leading $d_{x^2-y^2}$-wave channels relative to the $s_\pm$-wave ground state $\lambda_{d(i)}\lambda_s$ ($s$, $d(1)$, and $d(2)$ correspond to $\alpha=1,~2$, and 3, respectively) is predicted to increase with doping (full symbols) in qualitative agreement with experiment (open symbols). The results on the basis of fRG and RPA are similar. From \cite{Bohm:2018} with permission.
  }
  \label{fig:RPA-fRG}
\end{figure}

In contrast to the Hubbard-Holstein model, which predicts an $s_{++}$ state \cite{Kontani:2010}, both RPA and fRG find an $s_\pm$ ground state for the specific band structure of the FeBCs where the energy gap has the same magnitude on the electron and hole bands but opposite sign \cite{Mazin:2008,Thomale:2011,Thomale:2009,Chubukov:2008b,Platt:2014}. Upon using a realistic band structure \cite{Graser:2009} and a rigid band model for simulating the doping, the hierarchy of pairing interactions was studied with RPA and fRG schemes. The results are similar in both cases and show that the ground state is $s_\pm$ wave followed by two $d_{x^2-y^2}$ pairing tendencies. The solution of the eigenvalue equations yields the eigenvectors $g_\alpha({\bf k})$ and eigenvalues $\lambda_\alpha$ in channel $\alpha$. $g_\alpha({\bf k})$ and $\lambda_\alpha$ describe the variation with {\bf k} of the energy gap and, respectively, the coupling strength in channel $\alpha$. On this basis the positions of the BS modes can be predicted and compared with the experiments as shown in Fig.~\ref{fig:RPA-fRG}.

The agreement of experiment and theory is remarkable in the case of fRG and still qualitative for RPA. The RPA results are offset to lower coupling strengths by 10 to 20\%. Since RPA neglects contributions other than spin-fluctuations, the observed discrepancy may indicate the existence of weak contributions from other coupling mechanisms such as charge fluctuations \cite{Bohm:2018}. Yet, the similarity of fRG and RPA results for the hierarchy of pairing channels supports spin-fluctuation induced superconductivity in BKFA in the doping range studied. CKFA appears to fit into this picture although the weakness of the putative low-energy BS mode, the proximity of the second BS mode to the pair-breaking maximum \cite{Jost:2018}, and the resulting controversy in the interpretation \cite{ZhangWL:2018} qualify this conclusion and call for further experiments.

Whereas the recent RPA and fRG studies favor spin fluctuations, an $s_\pm$-wave gap and two sub-leading pairing channels having $d_{x^2-y^2}$ symmetry the Hubbard-Holstein model leads to different conclusions and finds an $s_{++}$ ground state driven by electron-phonon coupling and orbital fluctuations \cite{Kontani:2010}. For the construction of the model sub-leading channels were not identified. However, if the sub-leading interactions in this model would be identified to have $d_{x^2-y^2}$ symmetry the resulting Raman spectra would be indistinguishable from those observed in BKFA and CKFA. Thus the case for spin fluctuation induced pairing depends crucially on the reliability of the hierarchy of pairing tendencies derived from fRG and RPA. Other experimental probes such as the study of quasi-particle interference effects in magnetic fields \cite{Hanaguri:2010} or in the presence of impurities \cite{Hischfeld:2015,Boker:2019} by scanning tunneling spectroscopy may help to clarify the symmetry of the ground state.

\section{Conclusions}
Raman scattering in iron pnictides and chalcogenides has provided a host of information on the electronic, magnetic and lattice properties of these systems. We focused on the the spin and charge degrees of freedom in this review.

In all cases the spectra consist of a superposition of several types of excitations. To which extent \textit{luminescence} (as an \textit{a priori} undesired contribution) plays a role is not entirely clear, but the comparison of a large amount of results shows that luminescence decreases substantially with improved sample quality and may be neglected at least at low energies.

\textit{Particle-hole excitations} are important in all compounds and for all doping levels. They are partially gapped out in the SDW state (section \ref{sec:SDW}) where the materials remain metallic and fully gapped out in superconducting BKFA, for instance, when the surfaces are sufficiently clean (see section \ref{sec:gap spectroscopy} and Ref.~\cite{Kretzschmar:2013}). In the normal state, the particle-hole excitations in \Alg symmetry depend on temperature as expected from the static resistivity [cf. Fig.~\ref{fig:Kretzschmar}\,(d)].

In \Blg symmetry, a strong contribution from \textit{fluctuations} (see section \ref{sec:fluct-exp}) is observed below room temperature at energies of order $k_{\rm B}T$ which softens with decreasing temperature, has the strongest spectral weight directly at the structural transition \Ts, and loses intensity below \Ts without, however vanishing so long as the material does not order magnetically. Raman scattering is particularly useful here since other spectroscopies have generic difficulties in observing the fluctuations: In the case of neutron scattering, the fluctuations appear only in the notoriously weak four-particle correlation, NMR spectroscopy cover only a small energy range well below $k_{\rm B}T$. The same holds true for thermodynamic methods \cite{Bohmer:2016} or transport \cite{Chu:2012} both of which probe fluctuations only indirectly. We argue that the temperature dependence identifies the Raman response as critical fluctuations which are expected to vanish at the related transition temperature. The persistence of the excitation below \Ts is, therefore, considered an indication of spin rather than charge fluctuations. Yet, there is no consensus in the published literature on this point. However, if this controversy could be settled the driving force behind the phase transitions would be identified.

In all pnictides, the response of the \textit{SDW} is clearly observed (section \ref{sec:SDW}). Although the gap energies are in the range 6-8\,$k_{\rm B}\TSDW$, the spectra are by and large described by weak-coupling physics including band reconstruction in the ordered state. This indicates that the magnetism here results from a Fermi surface instability of itinerant electrons. The relevant gap energies are in the range 100-150\,meV in the parent compounds and decrease with doping or substitution along with the gradual suppression of the magnetically ordered phase.

In the chalcogenide Fe(Se$_{1-x}$S$_x$) the \Blg response is distinctly different at all temperatures \cite{Zhang:2017s,Baum:2019}. In the range 60\,meV, a broad excitation is observed for $x<0.2$ which gains spectral weight upon cooling by a factor of approximately two without moving by more than a few percent. This temperature (and doping) dependence is not expected for quasi-particle scattering from impurities. Rather, the peak was associated with \textit{two-magnon excitations in a frustrated magnet} (see sections \ref{sec:local-spin} and \ref{sec:origin}) of nearly localized moments in agreement with neutron scattering experiments, LDA predictions for the exchange parameters $J_1$ and $J_2$, and simulations using exact diagonalization \cite{Baum:2019,Ruiz:2019}. The response from fluctuations entirely fills the gap below 60\,meV in the temperature range around \Ts and persists down to $T\approx 20$\,K. Indications of an SDW were not found. It is argued that orbital dependent localization of electrons as expected in Hund's metals with $J\sim U$ may be at the origin of this dichotomy between the pnictides and chalcogenides \cite{Yin:2011,Si:2016}.

In the superconducting state \textit{gap excitations} are observed in all sufficiently clean systems independent of the concentration of substitutional atoms (see section \ref{sec:gap spectroscopy}). Only Co substitution gradually suppresses the pair-breaking features. In all FeBCs there is a strong band dependence of the gaps. In FeSe the gap can be resolved in the Raman spectra but its small magnitude  prevents a reliable analysis. In BFCA, the gaps on the electron bands exhibit a strong modulation with momentum and may even have accidental nodes at optimal doping. Here, Raman scattering and transport measurements arrive at similar conclusions \cite{Muschler:2009,Mazin:2010a,Tanatar:2010a}. In BKFA and presumably CKFA, the gaps on the individual bands are nearly constant. This fact, first derived from ARPES \cite{Evtushinsky:2009,Mou:2016}, manifests itself in sharp gap edges in the Raman spectra. 

Below the gap edges narrow, nearly \textit{resolution-limited lines} are observed in the \Blg spectra \cite{Bohm:2014} (see sections \ref{sec:collective} and \ref{sec:doping}). These lines display a BCS-like temperature dependence, vary as $1-x$ with doping and steal spectral weight from the pair-breaking features. The pair-breaking features scale with \Tc and barely depend on temperature. These criteria are predicted only for BS modes that result from sub-leading pairing interactions competing with the ground state. 

\textit{Microscopic model calculations} using fRG and RPA (see section \ref{sec:pairing}) show that the pnictides have indeed a hierarchy of pairing channels with very similar eigenvalues, an $s_\pm$ ground state, and two sub-leading $d_{x^2-y^2}$ instabilities of different order \cite{Bohm:2018}. The Raman experiments agree semi-quantitatively with these predictions, in particular with the symmetry, but cannot pin down the sign change of the ground state. Tunneling experiments in samples with different impurity concentration and with applied field may settle this point. Yet, the doping dependence of the sub-leading channels in BKFA and presumably the results in CKFA as well make a strong case for spin fluctuations to contribute partially or predominantly to the Cooper pairing in the pnictides.

In summary, the most significant contributions from light scattering experiments to the physics of the FeBCs pertain to the analysis of fluctuations and of the superconducting pairing states. The fluctuations can be compared to the evolution of the elasticity \cite{Gallais:2016a,Bohm:2016} and of the spin-lattice relaxation as obtained from NMR studies \cite{Thorsmolle:2016}. While the interrelation of the various methods is obvious several aspects of the interpretation remain controversial, in particular the origin of the fluctuations. Concerning superconductivity the derivation of the gap energies is of specific relevance. Table~\ref{tab:gap} shows that the results from light scattering fit very well into the concert of the other methods if the data are read properly. Reading properly means, in particular, understanding the respective observables and including collective excitations which reveal details of the pairing potential $V_{{\bf k},{\bf k}^\prime}$. In many cases the Raman response contributes information which cannot easily or not at all be obtained by other methods. Thus, part of the understanding of the pnictides and chalcogenides may rest on light scattering results, in particular if the remaining challenges in the interpretation can be settled and if experiments under extreme conditions, specifically pressure, can be applied more routinely in the future.\\

\section*{Acknowledgements}
We gratefully acknowledge discussions with Istv\'an T\"utt\H{o}, A. Baum, A. Chubukov, S. Maiti, F. Hardy, C. Meingast, P. Hirschfeld, B. Moritz, T.P. Devereaux, D. Jost, Z.V. Popovi\'c, W. Hanke, and R. Thomale.

Financial support came from the German Research Foundation (DFG) via the Priority
Program SPP\,1458 (grant-no. Ha2071/7) and the Transregional Collaborative Research Center TRR80 and by the Serbian Ministry of Education, Science and Technological Development under {Project III45018}. We acknowledge support by the DAAD through the bilateral project between Serbia and Germany (grant numbers  57335339 and 57449106).\\\\

\bibliographystyle{iopart-num}
\bibliography{literatureR2}

\end{document}